\documentclass[iop]{emulateapj}
\pdfoutput=1

\usepackage{graphics, color}

\newcommand{\myvec}[1]{\mathbf{#1}}

\def\subsun{\mbox{$_{\odot}$}}
\def\rnorm{\mbox{$r/r_{200}$}}
\def\rvir{\mbox{$r_{200}$}}
\def\fsp{\mbox{$f_{sp}$}}

\shorttitle{Simulations of Conductive Clusters}
\shortauthors{Smith et al.}

\begin{document}

\title{Cosmological Simulations of Isotropic Conduction in Galaxy Clusters}

\author{Britton Smith, 
  Brian W. O'Shea\altaffilmark{1},
  G. Mark Voit,
  David Ventimiglia}
\affil{Department of Physics \& Astronomy,
  Michigan State University,
  East Lansing, MI 48824}

\and

\author{Samuel W. Skillman\altaffilmark{2}}
\affil{Center for Astrophysics and Space Astronomy, 
  Department of Astrophysical \& Planetary Science, 
  University of Colorado, 
  Boulder, CO 80309, USA}

\altaffiltext{1}{Lyman Briggs College and Institute for Cyber-Enabled Research, Michigan State University, East Lansing, MI 48824, USA}

\altaffiltext{2}{DOE Computational Science Graduate Fellow}

\email{smit1685@msu.edu}

\begin{abstract}
Simulations of galaxy clusters have a difficult time reproducing
the radial gas-property gradients and red central galaxies observed 
to exist in the cores of galaxy clusters.  Thermal conduction has been suggested as a 
mechanism that can help bring simulations of cluster cores into better alignment with
observations by stabilizing the feedback processes that regulate gas
cooling, but this idea has not yet been well tested with cosmological numerical
simulations.  Here we present cosmological simulations of ten galaxy clusters
performed with five different levels of isotropic Spitzer conduction, which 
alters both the cores and outskirts of clusters, but not dramatically.  In the cores, 
conduction flattens central temperature gradients, making them nearly isothermal and 
slightly lowering the central density but failing to prevent a cooling catastrophe there.  
Conduction has little effect on temperature gradients outside of
cluster cores because
outward conductive heat flow tends to inflate the outer parts of the intracluster 
medium (ICM) instead of raising its temperature.  
In general, conduction tends reduce temperature inhomogeneity in the ICM, 
but our simulations indicate that those homogenizing effects 
would be extremely difficult to observe in $\sim 5$~keV clusters.  
Outside the virial radius, our conduction implementation lowers the gas densities 
and temperatures because it reduces the Mach numbers of accretion shocks.  
We conclude that despite the numerous small ways in which conduction alters 
the structure of galaxy clusters, none of these effects are significant enough 
to make the efficiency of conduction easily measurable unless its effects are
more pronounced in clusters hotter than those we have simulated.
\end{abstract}


\section{Introduction}

Numerical simulations of galaxy clusters have not yet succeeded in
producing objects with properties identical to those of observed
galaxy clusters.  The most serious discrepancies with observations are
in the cores \citep[e.g,][]{2007ApJ...668....1N, 2011ASL.....4..204B, 2013ApJ...763...38S}.
Within the central $\sim$100 kpc of clusters whose central cooling
time is less than a Hubble time, radiative cooling of intracluster gas
in the simulations tends to be too efficient, leading to
overproduction of young stars and excessive entropy levels in the core
gas, as higher-entropy material flows inward to replace the gas that
has condensed.  The resulting cluster cores in such simulations consequently 
have temperature profiles that decline from $\sim$10 kpc outward, in stark
disagreement with the observed temperature profiles of cool-core
clusters, which rise in the $\sim$10--100 kpc range.

Feedback from a central active galactic nucleus helps solve this
problem because it slows the process of cooling.  However, the core
structures of clusters produced in cosmological simulations 
depend sensitively on the details of the implementation of feedback,
the radiative cooling scheme, and numerical resolution.  For example,
\citet{2011MNRAS.417.1853D} simulated the same cosmological galaxy
cluster with two different mechanisms for feedback and two different
cooling algorithms, one with metal-line cooling and one without.  The
core structures of the resulting clusters were quite diverse, with
kinetic jet feedback and metal-free cooling producing the most
realistic-looking cores.  But discouragingly, allowing cooling via
metal lines had the effect of supercharging AGN feedback and
producing a cluster core with excessively high entropy.

Another oft-proposed solution to the core-structure problem in galaxy
clusters is thermal conduction.  It has been invoked many times to
mitigate the prodigious mass flows predicted for cool-core clusters
lacking a central heat source \citep[e.g.,][]{1983ApJ...267..547T,
  1986ApJ...306L...1B, 2001ApJ...562L.129N} but cannot be in
stable balance with cooling \citep[e.g.,][]{1988ApJ...326..639B,
  2003MNRAS.342..463S, 2008ApJ...688..859G} and does not completely
compensate for radiative cooling in all cluster cores
\citep{2002MNRAS.335L...7V, 2003ApJ...582..162Z}.  Nevertheless, the
temperature and density profiles of many cool-core clusters are
tantalizingly close to being in conductive balance, suggesting that
AGN feedback is triggered only when inward thermal conduction, perhaps
assisted by turbulent heat transport, fails to compensate for radiative losses
from the core \citep[e.g.,][]{2011ApJ...740...28V}.

The nebular line emission observed in many cool-core clusters also
suggests that thermal conduction may be important.  Ultraviolet light
from young stars accounts for some of the line emission but cannot
explain all of it, implying that another heat source is present
\citep[e.g.,][]{1987MNRAS.224...75J, 1997ApJ...486..242V, 2013ApJ...767..153W}.  
Observations of the optical and infrared line ratios are
consistent with heat input by a population of suprathermal electrons,
possibly entering the nebular gas from the ambient
hot medium \citep{2009MNRAS.392.1475F}.  
The emission-line studies are not conclusive on this
point, but recent observations of ultraviolet emission from a filament
in the Virgo cluster have bolstered the case for conduction
\citep{2009ApJ...704L..20S, Sparks+11}.  The C~IV 1550~\AA\ and He~II
1640~\AA\ emission lines from this filament are consistent with a
model in which the filament is surrounded by a conductive sheath that
channels heat into the filament.

If conduction is indeed important in determining the structure of
cluster cores and mediating the heat flow into cool gas clouds, then
it ought to be included in cosmological simulations of galaxy
clusters.  \citet{2004ApJ...606L..97D} presented the first simulations
of conductive intracluster media in cosmological galaxy clusters,
using the \citet{2004MNRAS.351..423J} implementation of isotropic
conduction in the \texttt{GADGET} smooth-particle hydrodynamics code.  These
simulations were notable in that they vividly demonstrated the
inability of conduction to prevent a cooling catastrophe.  Even at 1/3 of 
the full Spitzer rate, thermal conduction could not prevent a large
cooling flow from developing in the cluster core.  Conduction delayed
the cooling catastrophe but did not diminish the magnitude of the
eventual mass inflow.  An updated set of cluster simulations using the
same conduction algorithm was recently performed by
\citet{2011MNRAS.416..801F}.  Isotropic conductivity was set to 1/3 of
the full Spitzer value for simulations including cooling and star
formation but without AGN feedback.  As with the previous simulation
set, there was not much difference between the global properties of
the final states of clusters simulated with and without thermal
conduction.

Conduction is also expected to smooth out temperature inhomogeneities 
in hot clusters \citep[e.g.,][]{2003ApJ...586L..19M}.  
One of the clusters simulated by \citet{2004ApJ...606L..97D} 
was quite hot, with a peak gas temperature of $\sim 12$~keV, and conduction in that
cluster produced a much more homogeneous ICM, with far less azimuthal
temperature variation than its conduction-free counterpart.  
Producing simulated clusters with small-scale temperature variations
similar to those of observed clusters continues to be somewhat tricky 
\citep[see, for example,][]{2012ApJ...747..123V}.   In particular, some 
recent simulations of sloshing, magnetized cluster cores with anisotropic 
conduction are suggesting that conduction needs to be somewhat suppressed
{\em along} the magnetic field lines into order to explain observations of cluster
cores with cold fronts \citep{2013ApJ...762...69Z}

Implementations of anisotropic conduction directed along magnetic
field lines are also of interest because the MHD instabilities that
result may have important implications for the structure of cluster
cores and the generation of turbulence in cluster outskirts 
\citep[e.g.,][]{2009ApJ...703...96P,2012MNRAS.419L..29P}.
\citet{2011ApJ...740...81R} recently simulated a cosmological cluster
with anisotropic conduction using the \texttt{FLASH} adaptive
mesh-refinement (AMR)
code.  Runs were performed with radiative cooling on and off, but no
AGN feedback was included.  In this case, anisotropic conduction
significantly reduced mass accretion into the central galaxy, even
though the magnetic field geometry significantly suppressed radial
heat conduction.  However, the $\sim 31$ kpc/$h$ 
effective resolution was insufficient to adequately resolve core
structure.

In this paper we present a new set of cosmological cluster simulations
performed with the AMR code \texttt{Enzo}, including isotropic
Spitzer conduction.  We simulate 10 clusters spanning a mass range 
of $2 - 8 \times 10^{14} \, M_\odot$.  We also model a range of suppression factors in order to
evaluate the effects of thermal conductivity on core structure.  As
with previous calculations, we do not include AGN feedback, and a
cooling catastrophe therefore results, once again confirming that
conduction alone cannot prevent strong cooling flows from developing
in galaxy clusters.  However, our focus here is on how conduction affects
the radial profiles and homogeneity of gas density and temperature outside 
the central $\sim 20$~kpc.

Our presentation of these conductive cluster simulations proceeds as
follows.  In \S~2 we describe the primary features of the
\texttt{Enzo} code employed in this work and describe our recent
implementation of isotropic Spitzer conduction, with the simulation
setup given in \S~3.  Section~4 discusses the qualitative morphological 
properties of the simulated clusters, while \S~5 compares their radial profiles.
Section~6 focuses on conduction affects the thermal homogeneity of the ICM,
\S~7 looks at the star formation histories of the clusters, and \S~8 summarizes 
our results.

\section{Numerical methods}

In this work, we use the open-source cosmological, adaptive
mesh-refinement (AMR) hydrodynamic + N-body code
\texttt{Enzo}\footnote{http://enzo-project.org/}
\citep{bryan97,bryan99,norman99,oshea04, 2005ApJS..160....1O}.  Below,
we describe the primary features of \texttt{Enzo} used in this work, including
newly-added machinery for solving thermal conduction.  
All analysis is performed with the open-source simulation analysis
toolkit, \texttt{yt}\footnote{http://yt-project.org/}
\citep{2011ApJS..192....9T} using the Parallel-HOP halo finder
\citep{2010ApJS..191...43S}.

\subsection{The Enzo code} \label{sec:enzo}

The \texttt{Enzo} code couples an N-body particle-mesh (PM) solver
\citep{Efstathiou85, Hockney88} used to follow the evolution of a
collisionless dark matter component with an Eulerian AMR method for
ideal gas dynamics by \citet{Berger89}, which allows high dynamic
range in gravitational physics and hydrodynamics in an expanding
universe.  This AMR method (referred to as \textit{structured} AMR)
utilizes an adaptive hierarchy of grid patches at varying levels of
resolution.  Each rectangular grid patch (referred to as a ``grid'')
covers some region of space in its \textit{parent grid} which requires
higher resolution, and can itself become the parent grid to an even
more highly resolved \textit{child grid}.  \texttt{Enzo}'s implementation of
structured AMR places no fundamental restrictions on the number of
grids at a given level of refinement, or on the number of levels of
refinement.  However, owing to limited computational resources it is
practical to institute a maximum level of refinement, $\ell_{max}$.
Additionally, the \texttt{Enzo} AMR implementation allows arbitrary integer
ratios of parent and child grid resolution, though in general for
cosmological simulations (including the work described in this paper)
a refinement ratio of 2 is used.

Multiple hydrodynamic methods are implemented in \texttt{Enzo}.  For this work,
we use the method from the ZEUS code \citep{stone92a,stone92b} for its
robustness in conditions with steep internal energy gradients, which are common
in cosmological structure formation when radiative cooling is used.
We use the metallicity-dependent radiative cooling method described in
\citet{2008MNRAS.385.1443S} and \citet{2011ApJ...731....6S}.  This
method solves the non-equilibrium chemistry and cooling for atomic H
and He \citep{abel97,anninos97} and calculates the cooling and heating
from metals by interpolating from multi-dimensional tables created with
the photo-ionization software, \texttt{Cloudy}\footnote{http://nublado.org/} 
\citep{1998PASP..110..761F}.  Both the primordial and metal cooling
also take into account heating from a time-dependent, spatially
uniform metagalactic UV background \citep{2001cghr.confE..64H}.

To model the effects of star formation and feedback, we use a modified
version of the algorithm presented by \citet{1992ApJ...399L.113C}.  A
grid cell is capable of forming stars if the following criteria are
met: the baryon overdensity is above some threshold (here 1000), the
velocity divergence is negative (i.e., the flow is converging), the gas mass in the cell is greater
than the Jeans mass, and the cooling time is less than the self-gravitational dynamical
time.  Because conduction can potentially balance radiative cooling,
instead of the classical definition of the cooling time, $e/\dot{e}$,
we use a modified cooling time that includes the change in energy from
conduction, defined as
\begin{equation}
t_{cool} = \frac{e}{\dot{e}_{rad} + \dot{e}_{cond}},
\end{equation}
where $\dot{e}_{rad}$ is the cooling rate and $\dot{e}_{cond}$ is the conduction
rate.  In the original implementation, if all of the above criteria
are satisfied, then a star particle representing a large, coeval group of stars
with the following mass is formed:
\begin{equation} \label{eq:m_star}
  m_{*} = f_{*} \ m_{\rm cell} \ \frac{\Delta t}{t_{\rm dyn}} \; , 
\end{equation}
where $f_{*}$ is an efficiency parameter, $m_{\rm cell}$ is the baryon
mass in the cell, $t_{\rm dyn}$ is the dynamical time of the combined
baryon and dark matter fluid, and $\Delta t$ is
the timestep taken by the grid containing the cell in question.  The factor of $t_{dyn}$ is added to
provide a connection between the physical conditions of the gas and
the timescale over which star formation and feedback occur.  When a
star particle is created, feedback in the form of thermal energy, gas,
and metals is returned to the grid at a rate given by
\begin{equation} \label{eq:m_star_form}
  \Delta m_{\rm sf} = m_{*} \ \frac{\Delta t}{t_{\rm dyn}} \  \frac{(t - t_{*})}
  {t_{\rm dyn}}  \ e^{-(t - t_{*}) / t_{\rm dyn}}  \; , 
\end{equation}
where $t_{*}$ is the creation time for the particle, so that  $\Delta m_{\rm
  sf}$ rises linearly over one dynamical time, then falls off
exponentially after that.  This feature ensures that the feedback response
unfolds over a dynamical time, even if star formation in the simulation
is formally instantaneous.
We also use a distributed feedback model, in which
feedback material is distributed evenly over a $3\times3\times3$ cube
of cells centered on the star particle's location.  This was shown by
\citet{2011ApJ...731....6S} to more effectively transport hot,
metal-enriched gas out
of galaxies and to minimize overcooling issues that are common to
simulations of structure formation.  This method was also shown by
\citet{2013ApJ...763...38S} to provide a better match to observed
cluster properties than injecting all feedback into a single grid
cell.

In practice, \texttt{Enzo} simulations often implement an additional 
restriction, allowing creation of a star particle only if
$m_{*}$ is above some minimum mass.  This prevents the formation
of a large number of low-mass star particles whose presence can
significantly slow down a simulation.  However, the use of this
algorithm in concert with conduction must be done with care.  As we
discuss in \S~\ref{sec:conduct}, explicit modeling of 
thermal conduction can require timesteps that are significantly
shorter than the hydrodynamic Courant condition, which has the effect
of considerably reducing the value of $m_{*}$.  If the value of
$m_{*}$ is only slightly less than the minimum particle size, this
condition will likely be satisfied in a relatively short time as condensation
proceeds, causing the gas density to increase and its dynamical time 
to decrease.  When the conduction algorithm is active, the resulting 
time delay can allow cold, dense gas to persist while in thermal
contact with the hot ICM.  Since radiative cooling is proportional to
the square of the density, heat conducted into high-density gas can be
easily radiated away.  This artificial time delay therefore produces 
a spurious heat sink in the ICM that can potentially boost 
the rates of cooling and star formation.  In order
to avoid this unphysical behavior, we remove the factor of $\Delta
t/t_{dyn}$ from Equation \ref{eq:m_star} and adopt a constant value of
10 Myr for $t_{dyn}$ for use in Equation \ref{eq:m_star_form}.  We have
tested the effects of these changes by comparing two simulations run 
without conduction using both the original star-formation implementation 
and this modification and find the stellar masses at any given time 
differ by less than one percent.

\subsection{Conduction Algorithm}
\label{sec:conduct}

We implement the equations of isotropic heat conduction in a manner
similar to that of \citet{2004MNRAS.351..423J}, where the heat flux,
$j$, for a temperature field, $T$, is given by 
\begin{equation}
j = -\kappa \nabla T,
\end{equation}
where $\kappa$ is the conductivity coefficient.  In an ionized plasma, heat
transport is mediated by Coulomb interactions between electrons.  In
this scenario, referred to as Spitzer conduction
\citep{1962pfig.book.....S}, the conductivity coefficient is given by 

\begin{equation} \label{eq:kappa_sp}
  \kappa_{sp} = 1.31 n_{e} \lambda_{e} k \left( \frac{k_BT_{e}}{m_{e}} \right)^{1/2},
\end{equation}
where $m_{e}$, $n_{e}$, $\lambda_{e}$, and $T_{e}$ are the electron
mass, number density, mean free path, and temperature, and $k_B$ is
the Boltzmann constant.  The electron mean free path is
\begin{equation}
  \lambda_{e} =  \frac{3^{3/2}(k_BT)^{2}}{4\pi^{1/2}e^{4}n_{e} \ln \Lambda},
\end{equation}
where $e$ is the electron charge and $\ln \Lambda$ is the Coulomb
logarithm, defined to be the ratio of the maximum and minimum impact
parameters over which Coulomb collisions yield significant momentum
change in the interacting particles, which are electrons in this case.  
The Coulomb logarithm for electron-electron collisions is 

\begin{equation}
  \ln \Lambda = 23.5 - \ln\ (n_{e}^{1/2} T_{e}^{-5/4}) - \left( \frac{10^{-5} +
  (\ln\ (T_{e}) - 2)^{2}}{16} \right)^{1/2}
\end{equation}
\citep[][pg. 34]{Huba2011}.  Because of its relative insensitivity to
$n_{e}$ and $T_{e}$, we follow \citet{2004MNRAS.351..423J} and
\citet{1988xrec.book.....S} and adopt a constant value of $\ln \Lambda
= 37.8$, corresponding to $n_{e} = 1$ cm$^{-3}$ and $T_{e} = 10^{6}$
K.  For values more appropriate to the ICM ($n_{e} = 10^{-3}$
cm$^{-3}$, $T_{e} = 10^{7}$ K), $\ln \Lambda = 43.6$.  At a constant
density, $\ln \Lambda$ varies by $\sim$10-20\% over the range of
temperatures relevant here.  
The Spitzer conductivity then reduces to
\begin{equation}
  \kappa_{sp} = 4.9 \times 10^{-7}~T^{5/2}~{\rm erg~s^{-1}~cm^{-1}~K^{-1}}.
\end{equation}
Since $\ln \Lambda$ increases with $T$ and $\kappa_{sp} \propto 1 /
\ln \Lambda$, including a direct calculation of $\ln \Lambda$ would
serve to somewhat soften dependence of $\kappa_{sp}$ on $T$.
In low density plasmas with large temperature gradients, the
characteristic length scale of the temperature gradient, $\ell_T
\equiv T / |\nabla T|$, can be smaller than the electron mean free
path, at which point Equation \ref{eq:kappa_sp} no longer 
applies.  Instead, the maximum allowable heat flux is described by a
saturation term \citep{1977ApJ...211..135C}, given as 
\begin{equation}
   j_{sat} \simeq 0.4 n_e k_B T \left( \frac{2 k T}{\pi m_e} \right)^{1/2}.
\end{equation}
To smoothly connect the saturated and unsaturated regimes, we use an
effective conductivity \citep{1988xrec.book.....S} given by
\begin{equation}
  \kappa_{eff} = \frac{\kappa_{sp}}{1 + 4.2 \lambda_e / \ell_T}.
\end{equation}
The rate of change of the internal energy, $u$, due to conduction is then
\begin{equation} \label{eq:cond}
  \frac{du}{dt} = - \frac{1}{\rho} \nabla \cdot j.
\end{equation}

Our numerical implementation closely follows that of
\citet{2005ApJ...633..334P}.  We use an explicit, first-order, 
forward time, centered space algorithm, which is the most
straightforward to implement in an AMR code.  For each cell, Equation
\ref{eq:cond} is solved by summing the heat fluxes from all grid cell faces,
and calculating the electron density and temperature on the grid cell
face as the arithmetic mean of the cell and its neighbor sharing that
face.  The time step stability criterion for an explicit solution of
the conduction equation is
\begin{equation} \label{eq:t_cond}
  dt < 0.5 \frac{\Delta x^2}{\alpha}
  \label{eq-dt}
\end{equation}
where $\Delta x$ is the grid cell size and $\alpha$ is the thermal
diffusivity, defined as
\begin{equation}
  \alpha = \frac{\kappa}{n_e k_B}.
\end{equation}
Equation~(\ref{eq-dt}) has the potentially to be 
considerably more constraining than the hydrodyamical Courant
condition, which is proportional to $T^{-1/2}\Delta x $, and so the conduction 
timestep in conditions typical of the ICM is often much
shorter than the hydrodynamic timestep.  The conduction timestep is calculated
on a per-level basis and is taken to be the minimum of all such values
on a given level.  \texttt{Enzo} pads each AMR
grid patch with three rows of ghost zones from neighboring grids.
This allows the conduction routine to take three sub-cycled time steps for every
hydrodynamic step, and thus allows the minimum grid timestep to be a factor of
three larger than Equation \ref{eq:t_cond}.  After three steps,
temperature information from the outermost ghost zone has propagated
to the edge of the active region of the grid, so performing additional conduction
cycles would yield inconsistent solutions with neighboring grids.  By
default, \texttt{Enzo} rebuilds the adaptive mesh hierarchy for a
given refinement level after each timestep, which is computationally
expensive.  However, because conduction does not explicitly change the
density field (the value of which is the only quantity that determines
mesh refinement),
we modify this behavior such that the hierarchy is only rebuilt after
an amount of time has passed equivalent to what the minimum timestep
would be if conduction were not enabled.

This implementation models the case of isotropic Spitzer
conduction, where heat flows unimpeded along temperature gradients.  If
magnetic fields are present, and strong enough that the electron
gyroradius is small compared to the physical scales of interest (which
is likely to be true in the intracluster medium), then heat is restricted to flow 
primarily along magnetic field lines, so the heat flux becomes the dot product
of the temperature gradient with the magnetic field direction (i.e.,
$j = -\kappa \myvec{b} \myvec{b} \cdot \nabla T$, where $\myvec{b}$ is the
unit vector pointing in the direction of the magnetic field).
Diffusion perpendicular to magnetic field lines is generally
considered to be negligible.  We do not consider magnetic fields in
this work, but instead approximate the suppression of conduction by
magnetic fields by adding a suppression factor, $\fsp$, varying from 
0 to 1, to the heat flux calculation.  In reality, the strength and
orientation of magnetic fields in galaxy clusters is poorly
understood.  Hence, the degree to which conduction is suppressed from
its maximum efficiency is not known.  The extreme limit of tangled
magnetic fields is equivalent to isotropic conduction with $\fsp =
1/3$.  A number of works have shown that the presence of magnetic
fields alongside conduction can create a variety of instabilities that
greatly influence the effective level of isotropic heat transport 
\citep[e.g.,][]{2005ApJ...633..334P, 2007ApJ...664..135P,
  2010ApJ...713.1332R, 2011MNRAS.413.1295M, 2012MNRAS.419.3319M}.  For
this reason, we simulate a large range of values of $\fsp$, from
strongly suppressed ($\fsp = 0.01$) to fully
unsuppressed ($\fsp = 1$).

\section{Simulation Setup}

The simulations described in this work are are initialized at $z=99$
assuming the WMAP Year 7 ``best fit'' cosmological model
\citep{2011ApJS..192...16L, 2011ApJS..192...18K}: $\Omega_m = 0.268$,
$\Omega_b = 0.0441$, $\Omega_{CDM} = 0.2239$, $\Omega_\Lambda =
0.732$, $h=0.704$ (in units of 100 km/s/Mpc), $\sigma_8 = 0.809$, and
using an Eisenstein \& Hu power spectrum \citep{eishu99} with a
spectral index of $n = 0.96$.  We use a single cosmological
realization with a box size of 128 Mpc/$h$ (comoving) and a simulation resolution
of 256$^{3}$ root grid cells and dark matter particles.  From this
realization, we select the 10 most massive clusters and resimulate
each individually, allowing adaptive mesh refinement only in a region
that minimally encloses the initial positions of all particles that
end up in the halo in question at $z = 0$.  We allow a maximum of 5
levels of refinement, each by a factor of 2, refining on baryon and
dark matter overdensities of 8.  This gives us a maximum comoving spatial
resolution of 15.6 kpc/$h$.  We set the \texttt{Enzo}
parameter \texttt{MinimumMassForRefinementLevelExponent} to $-0.2$ for both
dark matter and baryon-based refinement, resulting in slightly 
super-Lagrangian refinement behavior.  For each halo in our sample, 
we perform a control simulation with $\fsp = 0$ and conductive simulations 
with $\fsp = 0.01$, 0.1, 0.3, and 1, evolving each to $z = 0$.  
All of the simulations employ the star
formation/feedback and radiative cooling methods described above in 
\S\ref{sec:enzo}.  Table \ref{tab:halos} lists the masses and
equivalent temperatures for all 10 clusters in the sample.

\begin{deluxetable}{lcc}
 \tablewidth{0pt}
  \tablecaption{Galaxy Cluster Sample}
  \tablehead{
    \colhead{ID} & \colhead{$M_{200}$ [$10^{14}\ M_{\odot}$]}
      & \colhead{$T_{200}$ [keV]}
       }
  \startdata
  0 & 8.04 & 5.56 \\
  1 & 8.26 & 5.66 \\
  2 & 5.60 & 4.37 \\
  3 & 5.56 & 4.35 \\
  4 & 3.76 & 3.35 \\
  5 & 4.64 & 3.85 \\
  6 & 3.85 & 3.40 \\
  7 & 3.52 & 3.20 \\
  8 & 1.91 & 2.13 \\
  9 & 3.41 & 3.14 \\
  \enddata
  \tablecomments{Masses and equivalent temperatures (as given by
    Equation \ref{eq:T200}) for the galaxy clusters in the sample.
    Note, the clusters are ordered by the masses computed by a halo
    finder in an original exploratory simulation.} \label{tab:halos}
\end{deluxetable}

\section{Qualitative Morphology}

\begin{figure}
  \centering
  \includegraphics[width=0.45\textwidth]{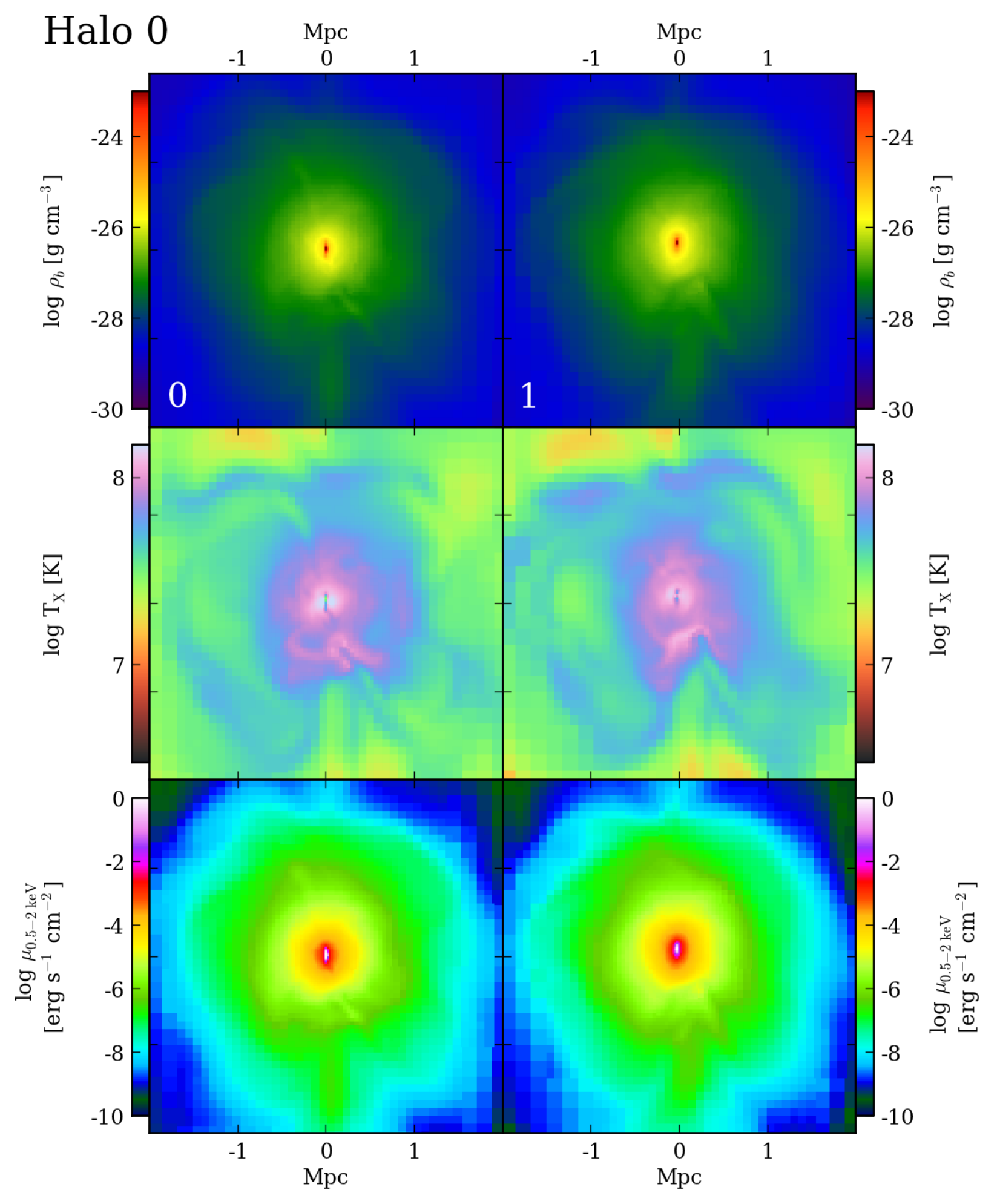}
  \caption{Projections of a 500 kpc thick slab centered on the most massive
    cluster at $z = 0$ 
    ($M_{200} = 8.0\times10^{14}\ M\subsun,\ T_{200} = 5.6$ keV) for the
    simulations with $\fsp = 0$ (left) and 1. (right)  The top and middle rows
    show average density and temperature, weighted by X-ray emissivity
    in the energy range of 0.5 to 2 keV.  The bottom row shows the
    X-ray surface brightness in the same energy range.
  } \label{fig:thin_proj_0}
\end{figure}

\begin{figure}
  \centering
  \includegraphics[width=0.45\textwidth]{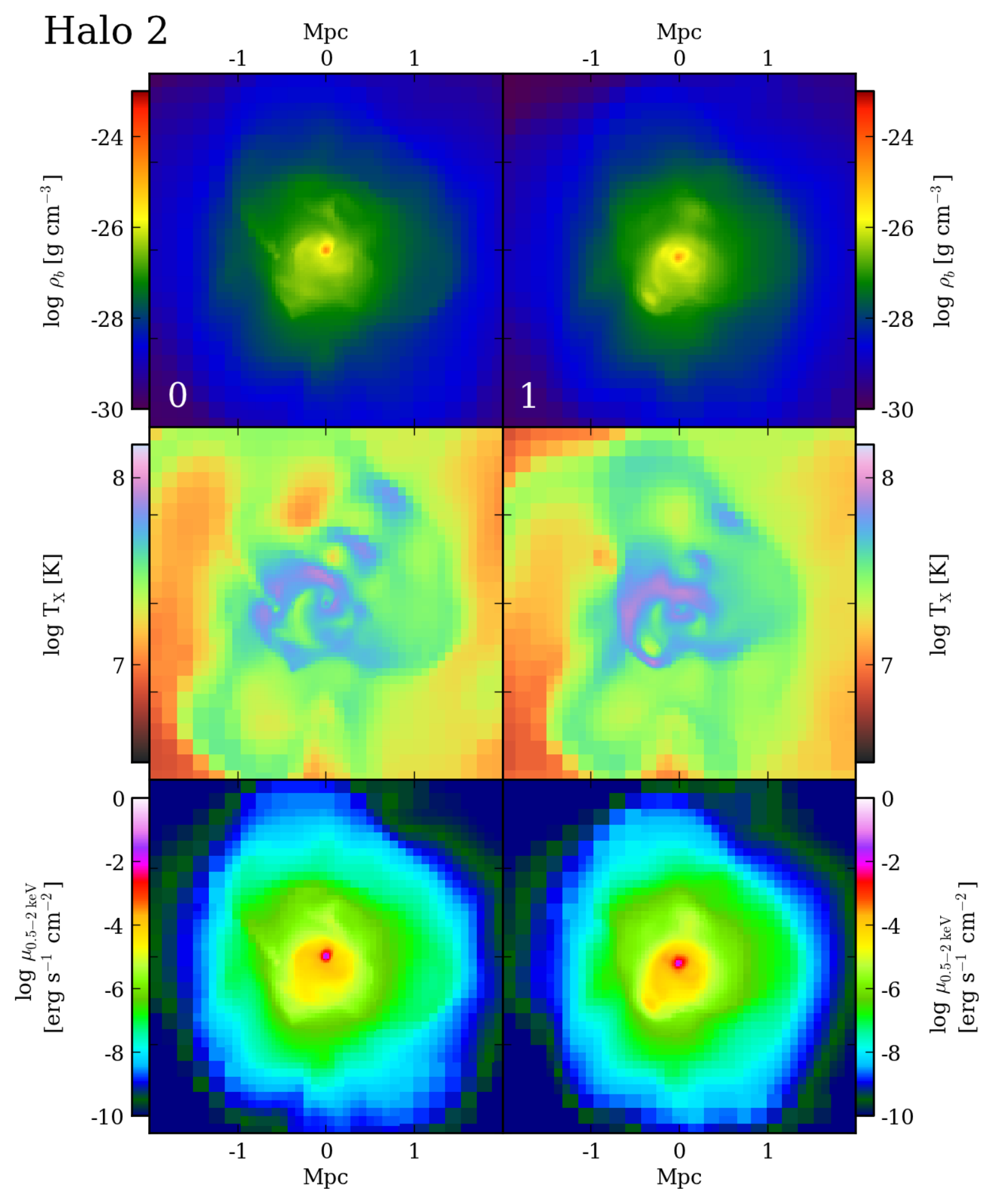}
  \caption{Projections of a 500 kpc thick  slab centered on the third most
    massive cluster at $z = 0$ 
    (M$_{200} = 5.6\times10^{14}\ M\subsun,\ T_{200} = 4.4$ keV) for
    the simulations with $\fsp = 0$ (left) and 1 (right).
  } \label{fig:thin_proj_2}
\end{figure}

\begin{figure}
  \centering
  \includegraphics[width=0.45\textwidth]{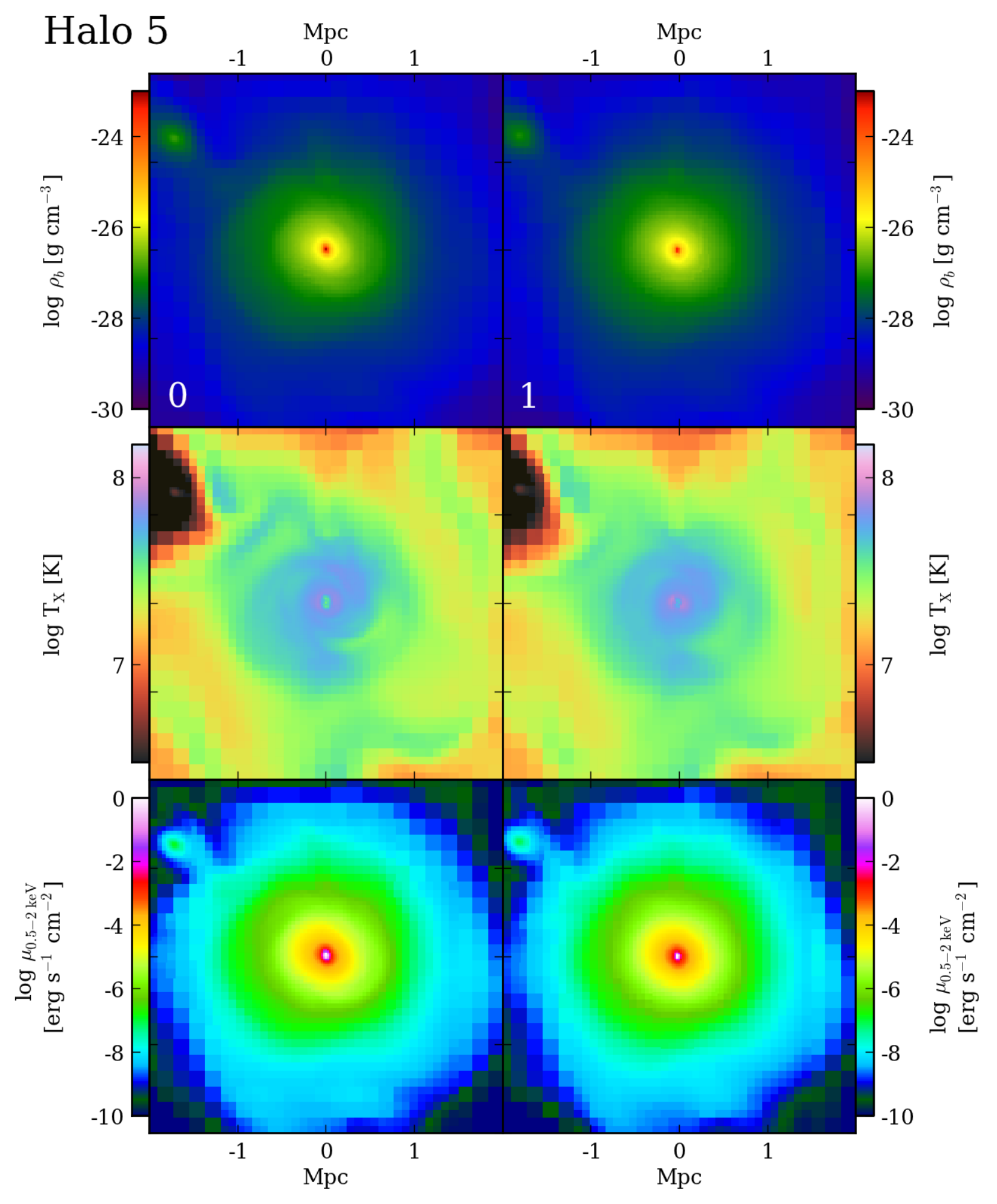}
  \caption{Projections of a 500 kpc thick  slab centered on the sixth most
    massive halo at $z = 0$ 
    (M$_{200} = 4.6\times10^{14}\ M\subsun,\ T_{200} = 3.9$ keV) for
    the simulations with $\fsp = 0$ (left) and 1 (right).
  } \label{fig:thin_proj_5}
\end{figure}

Perhaps the most surprising finding to emerge from our simulations is the
insensitivity of the qualitative morphological features of the ICM to thermal 
conduction, even in clusters with typical gas temperatures $\sim 5$~keV.
Figures \ref{fig:thin_proj_0}, \ref{fig:thin_proj_2}, and
\ref{fig:thin_proj_5} show gas density, temperature, and
X-ray surface brightness projections for three clusters at the minimum
and maximum conduction efficiency.  We choose these clusters in order to
capture a wide range of potential influence from conduction.  Halo 0
is the most massive cluster in the sample (with a mass of $8 \times
10^{14}$~M$_\odot$), Halo 2 is a highly 
unrelaxed cluster, and Halo 5 is relatively relaxed and is close to
the average mass for the sample (M$_{200}$ $\simeq 4.9 \times
10^{14}$~M$_\odot$)  In order to highlight differences
between values of $\fsp$, we limit the projected line of sight to 0.5
Mpc.  The morphological similarity between these conductive and
non-conductive clusters starkly contrasts with the obvious
differences found by \citet{2004ApJ...606L..97D} for a $\sim 12$~keV
cluster.  In their work, it is immediately obvious
which simulation included conduction.  

There are, however, some subtle but noticeable morphological differences 
between our conductive and non-conductive clusters.  The overall density
structure is slightly smoother for $\fsp = 1$ in all three clusters,
and the temperature field is visibly smoother in Halo 5.  In all three
clusters shown, the core appears to be marginally less dense with
conduction on and is also markedly hotter in Halos 2 and 5.
Nevertheless, while it is for the most part apparent that conduction
is present in our cluster sample with $\fsp = 1$, its effects are not nearly as
striking as in \citet{2004ApJ...606L..97D}.  Quite possibly, this difference 
is due to the strong temperature dependence of Spitzer conductivity, suggesting 
that it may be possible to estimate the typical conductivity of the ICM by analyzing
the dependence of azimuthal temperature homogeneity on cluster temperature 
(or lack thereof) in a large sample of galaxy clusters.

\section{Radial Profiles} \label{sec:profiles}

Systematic conductivity-dependent differences among our
simulated clusters are hard to see in individual cases but
are more apparent in comparisons of average properties of 
sample sets with differing conductivity.
Therefore, in order to understand the systematic effects of increasing the
conduction efficiency, we aggregate the radial profiles from all
the clusters in our simulated sample for each value of $\fsp$, as
shown in Figure \ref{fig:radial_profiles}.   To minimize artifacts that
result from the rebinning of profile data, we perform the initial
profiling in physical distance units to calculate the virial radius
for each cluster (which we take for convenience to be $r_{200}$
with respect to the critical density).
We then create a
second set of profiles in units of $\rnorm$ that are combined to make
the aggregate profiles presented here.  We refer to these as
``averaged profiles'' and employ this method for Figures
\ref{fig:radial_profiles} and \ref{fig:potential_profiles} and 
all subsequent figures showing aggregate properties. 

\subsection{Cluster Interiors} \label{sec:cluster_interiors}

The presence of conduction leads to higher gas temperatures 
both inside and outside of the conduction-free cluster's
characteristic temperature peak at $\rnorm \sim 0.07$.  
Within $\rnorm \sim 0.07$, gas temperatures in conductive clusters 
are greater than those in the non-conductive control simulation, with 
values of $\fsp \gtrsim 0.1$ leading to nearly isothermal cores and central
temperatures $\sim 40$\% greater than in non-conductive cluster simulations. 
The temperature profiles of our conductive clusters are in 
reasonably good agreement with those
of \citet{2011ApJ...740...81R} for their cluster with isotropic
conduction and $\fsp = 1/3$, although the core of their cluster is
less isothermal than ours, with cooler temperatures near the center.  
This could be because the simulations of 
\citet{2011ApJ...740...81R} did not include a prescription for star
formation and feedback in cluster galaxies, which can
increase the core temperature both by consuming the cold gas and
through thermal and mechanical feedback.  Agreement with the temperature 
profiles of the simulated clusters from \citet{2004ApJ...606L..97D} is not nearly as
good.  Those clusters show a significant reduction in both the peak
and central temperatures with conduction present, which we do not
observe.  The theoretical temperature profiles of
\citet{2013MNRAS.tmp.1104M} for the $10^{14.5} M\subsun$ clusters show an
elevation in the core temperatures in conductive clusters in rough agreement with our
results.  We also see a marginal inward movement of the location of
the temperature peak as they predict, although not nearly to the same
degree.  The temperature peak for our clusters is also at a
smaller radius to begin with.

\begin{figure*}
  \centering
  \includegraphics[width=1.0\textwidth]{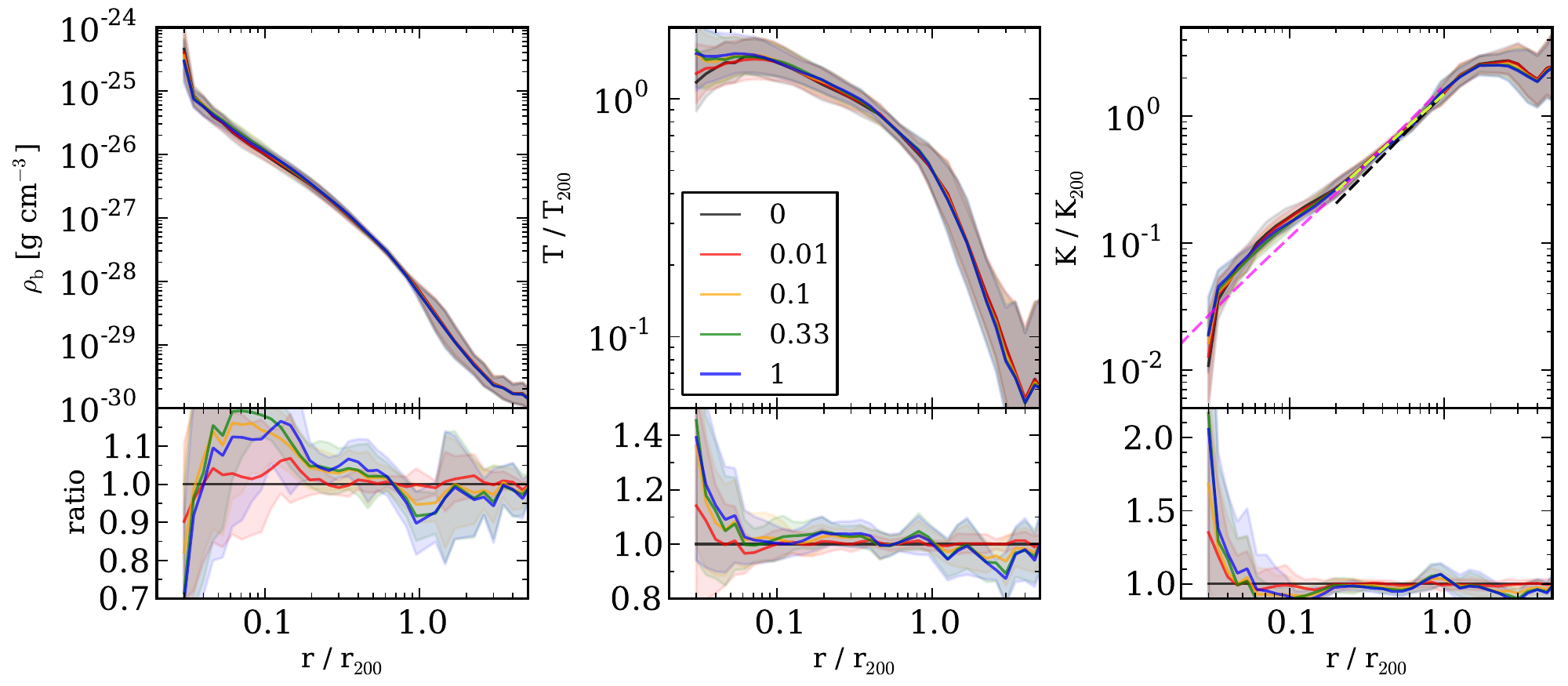}
  \caption{Averaged radial profiles for all ten halos in the sample,
    showing volume-weighted density, mass-weighted temperature, and
    entropy constructed from volume-weighted electron density and
    mass-weighted temperature profiles.  The colors denote the value
    of $\fsp$ and the shaded regions indicate the variance within
    the cluster sample.  The lower panels show profiles of the
    enhancement or decrement in the fields profiled above with respect
    to the clusters with simulated without conduction.  Note, the
    ratios shown in the lower panels were computed for each cluster,
    then aggregated, and as such are not strictly the ratio of the
    values plotted above.  The right panel also includes the power-law
    fit to non-radiative \texttt{Enzo} simulations of
    \citet{2005MNRAS.364..909V} (Equation \ref{eq:VKB_fit}, black,
    dashed line), a power-law fit to the sample presented here
    (Equation \ref{eq:our_fit}, dashed, lime-green line), and a pure
    cooling model for a 5 keV cluster \citep[][dashed, pink
      line]{2001Natur.414..425V, 2002ApJ...576..601V}.
  } \label{fig:radial_profiles}
\end{figure*}

Despite its ability to maintain approximate isothermality in
the cluster cores, conduction at even maximal efficiency is unable to
avert a cooling catastrophe at the very center of the halo.  This cooling
catastrophe and the resulting condensation and star-formation activity 
produce sharp peaks at $\rnorm \lesssim 0.03$ in all the gas density
profiles in Figure \ref{fig:radial_profiles}, as well as a slight temperature 
increase at the same location in the more highly conductive 
clusters.  The elevated central temperatures at the centers of conductive clusters
reduce the central gas density relative to the control clusters by
$\sim$20-30\%.  Material that would have fallen into the center is
displaced to larger radii, as can be seen by the elevated density at
$0.04 \lesssim \rnorm \lesssim 0.6$ in the more conductive runs.  
Gas density is enhanced primarily in the 
range $0.04 \lesssim \rnorm \lesssim 0.2$, but the enhancement
appears to saturate at $\sim$20\% for $\fsp = 0.33$.  For $\fsp =
1$, the density is actually lower than for $\fsp = 0.33$ in this
range, but is  then higher for $0.2 \lesssim \rnorm \lesssim 0.6$.  It is
unclear what causes the outward transport (or prevention of inflow) 
to stall just inside 0.2 $\rvir$, but it seems that it can be
overcome for some value of $\fsp$ above 0.33, likely closer to 1.

\begin{figure}
  \centering
  \includegraphics[width=0.45\textwidth]{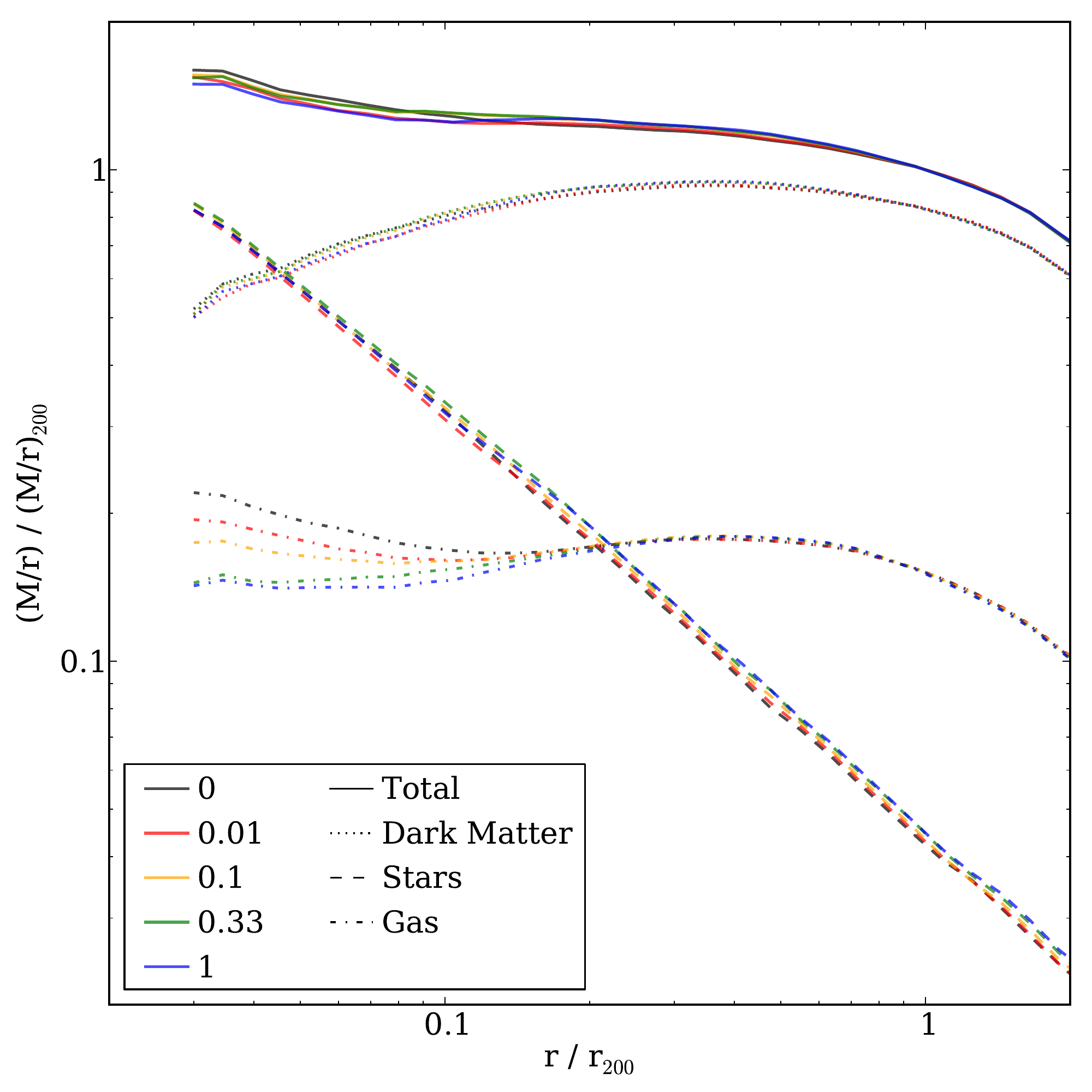}
  \caption{Averaged profiles of the radial mass distribution in terms of $M/r$, including
    individual contributions from dark matter, stars, and gas for all
    clusters.  The process by which the averaged profiles are created is
    described in \S\ref{sec:profiles} and in the caption of Figure
    \ref{fig:radial_profiles}.
  } \label{fig:potential_profiles}
\end{figure}

Further evidence of conduction-driven inflation of the core can be seen in Figure
\ref{fig:potential_profiles}, where we plot averaged profiles of 
$M/r$, including the individual contributions from
dark matter, stars, and gas.  Despite the lack of perfect
monotonicity in the density and temperature profiles, the gaseous
component of the potential decreases monotonically with increasing
$\fsp$, indicating that the clusters are indeed responding to the
presence of conduction by puffing up their cores and redistributing
gas to larger radii.  As is the case for most simulated galaxy
clusters, the stellar component is extremely centrally concentrated,
dominating the gravitational potential inside 0.04 $\rvir$.  The
stellar and dark matter components of the potential are largely
unaffected by conduction.  

In the right panel of Figure \ref{fig:radial_profiles}, we plot
normalized entropy profiles, where the normalization term, $K_{200}$
\citep{2005RvMP...77..207V,2005MNRAS.364..909V}, is given by
\begin{equation}
K_{200} = k_BT_{200}~\bar{n_{e}}^{-2/3},
\end{equation}
where $T_{200}$ is
\begin{equation} \label{eq:T200}
k_BT_{200} = \frac{G M_{200}~\mu m_{p}}{2 r_{200}},
\end{equation}
and $\bar{n_{e}}$ is the average electron number density within
$\rvir$, assuming a fully ionized plasma of primordial composition.
The decrease in density and increase in temperature 
in the cluster cores with conduction yield entropy profiles that are
enhanced by 40-70\%, but still show the steep decline in the very
center that is typical of simulated clusters.  This enhancement
decreases out to $\rnorm \sim 0.07$, where the mean entropy values are
approximately equivalent for all values of $\fsp$.  The entropy
profiles for the clusters with conduction then dip below the control
sample by roughly 10\% out to $\rnorm \sim 0.2$.  In the range $0.2
\lesssim \rnorm \lesssim 0.6$, the combined temperature and density
enhancements appear to be in perfect balance,  producing nearly
identical entropy profiles across the entire cluster sample with
a very small variance.

\citet{2005MNRAS.364..909V} find that for a sample of non-radiative,
non-conducting clusters simulated with \texttt{Enzo}, the entropy
profiles in the range $0.2 \lesssim \rnorm \lesssim 1$ are best fit by
the power-law
\begin{equation} \label{eq:VKB_fit}
K(r) / K_{200} = 1.51~(r / r_{200})^{1.24},
\end{equation}
which we overplot on Figure \ref{fig:radial_profiles} with a black
dashed line.  This power law matches our sample well in normalization,
but has a slightly steeper slope.  We find that our sample
is best matched by
\begin{equation} \label{eq:our_fit}
K(r) / K_{200} = 1.50~(r / r_{200})^{1.09}.
\end{equation}
For comparison, we also plot the predicted entropy profile for a pure
cooling model \citep{2001Natur.414..425V, 2002ApJ...576..601V} of a 5
keV cluster.  This model assumes gas and dark matter density
distributions that follow an NFW profile \citep{1997ApJ...490..493N}
with concentration $c = 6$.  The gas is initially in hydrostatic equilibrium and
allowed to cool for a Hubble time.  The pure cooling model agrees
quite well with our clusters in the range $0.2 \lesssim \rnorm
\lesssim 1$, and especially well in the range $0.2 \lesssim \rnorm
\lesssim 0.5$.  Outside of 0.5 $\rvir$, the slope of the pure-cooling
model is slightly too steep and is closer in slope to the
\citet{2005MNRAS.364..909V} fit.

\subsection{Cluster Exteriors}

The outskirts of galaxy clusters, while having lower temperatures and
hence lower conductivities, are not subject to intermittent heat
injection from stellar feedback, and thus offer an intriguing
laboratory for studying the effects of conduction.  We identify two
distinct regions in the cluster outskirts where conduction appears to
have influence, at $\sim \rvir$ and at $\sim3~\rvir$.  At $\rnorm
\gtrsim 0.6$, the density excess seen in the cluster interiors turns
into a deficit for all values of conduction simultaneously, as can be
see in Figure \ref{fig:radial_profiles}.  From this point
out to well past the virial radius, there exists a perfectly monotonic
trend of lower gas densities for higher values of $\fsp$.  For the maximum
value of $\fsp$, the average gas density at the virial radius is 10\%
lower than that of the clusters simulated without conduction.  In
fact, the density is measurably lower for all values of $\fsp \ge 0.1$
out to a few virial radii.  The temperature just inside $\rvir$ is
marginally higher, while the temperature just outside $\rvir$ is
reduced.  

Conduction transports heat outward in these regions because the gravitational
potential there produces a declining gas temperature gradient.  This heat transfer
causes the entropy of the outer gas to increase.  However, because
gas in the cluster outskirts is not pressure-confined like the gas in the core,  
it is free to expand outward and decrease in both density and temperature, while
its temperature gradient remains determined by the gravitational potential. 
This happens because the timescale for conduction in the outer regions is 
substantially greater than the sound-crossing time.  Consequently, conduction
of heat outward causes the outer gas to expand without much change in the
temperature gradient.  Beyond $\rvir$, however, we see a slight steepening 
in the temperature profile as predicted for conductive clusters
by \citet{2013MNRAS.tmp.1104M}, presumably because inflation of the ICM
due to conduction pushes gas near the virial radius farther from the cluster
center without adding much thermal energy.  The fact that the entropy profiles 
of conductive clusters are lower than those of non-conductive clusters beyond 
the virial radius supports this interpretation.

Further out, at $\sim3~\rvir$, there is another systematic decrease in
both density and temperature.  Interestingly, as was pointed out by
\citet{2008ApJ...689.1063S}, this is the typical location for a galaxy
cluster's accretion shocks, which are responsible for heating gas up to the
virial temperature.  This raises a critical question: 
\textit{How can conduction affect a large halo's accretion shocks?} 
 
\begin{figure*}
  \centering
  \includegraphics[width=1.0\textwidth]{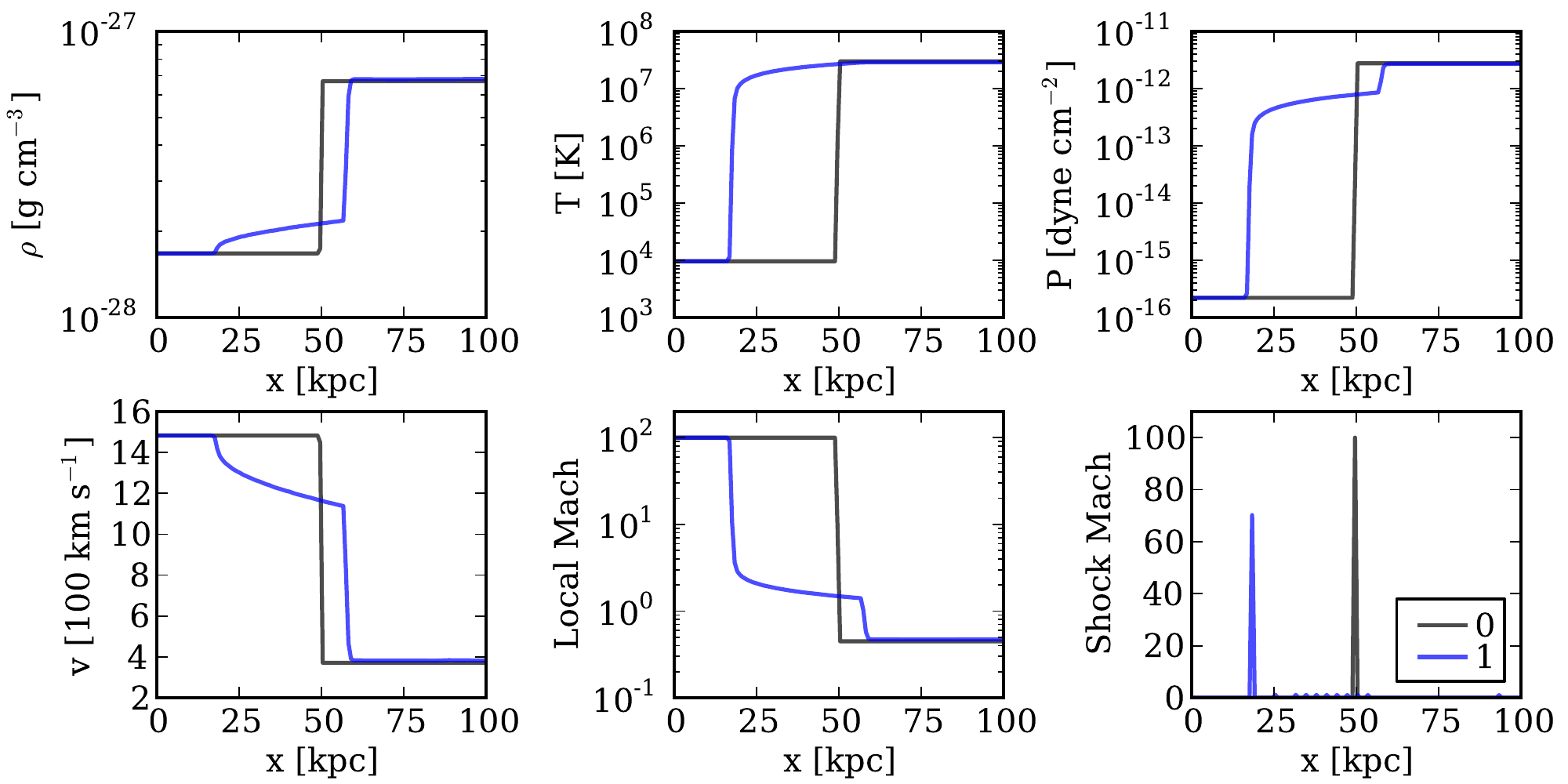}
  \caption{Profiles of a one-dimensional, Mach 100 standing shockwave
    after 200 Myr for $\fsp = 0$ and 1 (black and blue lines, respectively).  Note, the curves for $\fsp =
    0$ also denote the initial configuration of the shock, which is
    stationary in the case where conduction is negligible.
  } \label{fig:shocks_1d}
\end{figure*}

\begin{figure}
  \centering
  \includegraphics[width=0.45\textwidth]{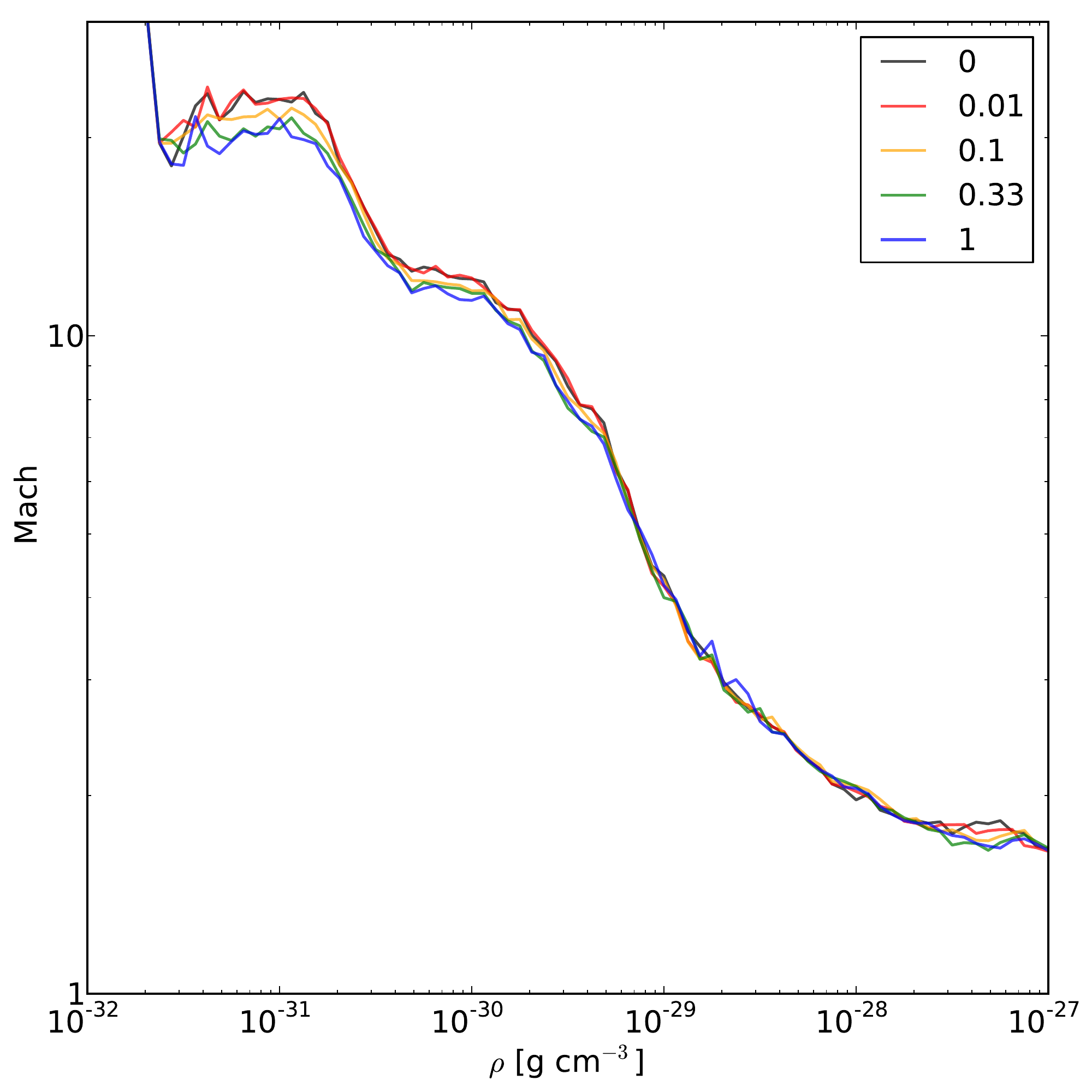}
  \caption{Profiles of the average Mach number as a function of
    gas density for the entire region where refinement is allowed in
    the simulations of the most massive galaxy cluster.  The average
    Mach number is calculated as a weighted mean, where the weight is
    the rate of kinetic energy processed through the shock, given by
    $\onehalf \rho v^{3} A$, where $\rho$, $v$, and $A$ are the
    pre-shock density, shock velocity, and shock surface area.
  } \label{fig:shock_profiles}
\end{figure}

To test this question in a more controlled environment
than a cosmological simulation, we
performed a pair of one-dimensional simulations designed to mimic the
conditions of an accretion shock around a galaxy cluster.  Following
the shock properties characterized by \citet{2008ApJ...689.1063S}, we
initiated a Mach 100 standing shockwave with preshock conditions of $n =
10^{-4}$ cm$^{-3}$ and $T = 10^{4}$ K, comparable to gas that is
falling directly onto a galaxy cluster (i.e., through spherical
accretion of the surrounding intergalactic medium, rather than being
accreted via filaments).  We ran simulations with $\fsp
= 0$ and 1, with profiles shown in Figure \ref{fig:shocks_1d}.  Within
tens of Myr, a conduction front in the $\fsp = 1$ simulation 
races ahead of the original shock.  The conduction front continues 
to advance with ever-decreasing speed and settles into a near-steady
state by $t = 200$ Myr, the time shown in the figure. 
Conduction of heat ahead of the main shock front therefore 
results in a separation of the density and temperature jumps, effectively 
creating two shocks, one produced by conductive preheating and 
a second nearly isothermal shock front some distance downstream.
The lower-right panel of Figure \ref{fig:shocks_1d} shows the 
Mach numbers determined using the shock-finding algorithm described 
in \citet{2008ApJ...689.1063S}.  Bifurcation of the original Mach 100 shock  
has produced a Mach $\sim 70$ conductive-precursor shock followed by 
a Mach $\sim 1.5$ shock where the main density jump occurs.
This splitting is also evident in the plot showing the local Mach ratio,
$v/c_{s}$ (lower-middle panel of Figure \ref{fig:shocks_1d}) on
either side of the temperature jump. The heat transfer upstream also
causes a slight drop in pressure at the original shock front, shifting
the primary density jump downstream by about 5 kpc.   

These results are independent of resolution.  The simulations shown in
Figure \ref{fig:shocks_1d} are for a grid 128 cells across, and we
observe nearly identical behavior down to a resolution of 16 cells.
For this configuration, the results are strongly dependent on $\fsp$.  At $\fsp =
0.67$, the distance between the separated shocks is approximately half
of that at $\fsp = 1$ and the Mach number of the primary shock is only
reduced to just under 90.  For $\fsp \lesssim 0.4$, the results are
indistinguishable from those without conduction.  However, we find
that it is possible to produce significant shockwave alteration for
lower values of $\fsp$ simply by lowering the initial Mach number and
increasing the preshock temperature (resulting in gas in a thermodynamic
regime comparable to gas that is being accreted from cosmological filaments).  As long as the postshock gas is
able to reach temperatures in the range of 10$^{7}$ K, where the
Spitzer conductivity becomes considerable, conduction is capable
of bifurcating the shock front.  

Figure~\ref{fig:shock_profiles} shows the average Mach number as a function
of gas density in our simulations of the most massive halo for all shocks in 
the subvolume in which refinement is allowed.  Conduction reduces the 
Mach numbers of the strongest shocks, which occur preferentially in the lowest 
density gas, by roughly 10\% from $\fsp = 0$ to $\fsp = 1$.  We therefore
conclude that the density, temperature, and entropy deficits observed 
beyond the virial radius in our conductive clusters are indeed due to shock
bifurcations similar to those seen in our idealized one-dimensional 
simulations of conductive shock fronts.  

Yet, the physics of the actual accretion shocks around real galaxy 
clusters is undoubtedly more complex.  
In particular, it is important to note that the electron mean free 
path in that gas is several times greater than the intershock distance in 
our one-dimensional simulations, and furthermore that real accretion
shocks are likely to be collisionless and magnetically-mediated. 
Nonetheless, the general qualitative point these simulations illustrate is
interesting:  Heating of preshock gas by a hot electron precursor has 
the potential to alter the expected relationships between the sizes 
and locations of the density and temperature jumps in accretion 
and merger shocks. 

Progress in understanding the effects of conduction on accretion shocks 
will require modeling the gas as a fully ionized plasma \citep{Zeldovich1957, 
Shafranov1957}, which is beyond the scope of this work.  
The high Mach number of an accretion shock combined with a post-shock 
temperature high enough for significant heat flux create conditions
similar to a supercritical radiative shock, as described by
\citet{2007ShWav..16..445L}.  When simulated with a two-fluid
approach, the combination of preheating of the preshock medium via
conduction in the electrons, electron-ion coupling, and compression of
the ion fluid in the postshock region can produce a small region,
known as a Zel'dovich spike, where the ion temperature is slightly
higher than the equilibrium postshock temperature
\citep{2007ShWav..16..445L, 2008ShWav..18..129L,
2011ShWav..21..367M}.   Our single-fluid simulations, despite
their limitations, produce shocks quite similar to the non-equilibrium 
results of \citet{2008ShWav..18..129L},
where a diffusion term proportional to $T^{5/2}$ (like Spitzer
conductivity) is used.  Nevertheless, a more detailed study of the 
characteristics of accretion shocks employing a two-fluid MHD treatment 
along with physically motivated conduction and
cooling rates seems warranted.

\section{Temperature Homogeneity}

\begin{figure}
  \centering
  \includegraphics[width=0.45\textwidth]{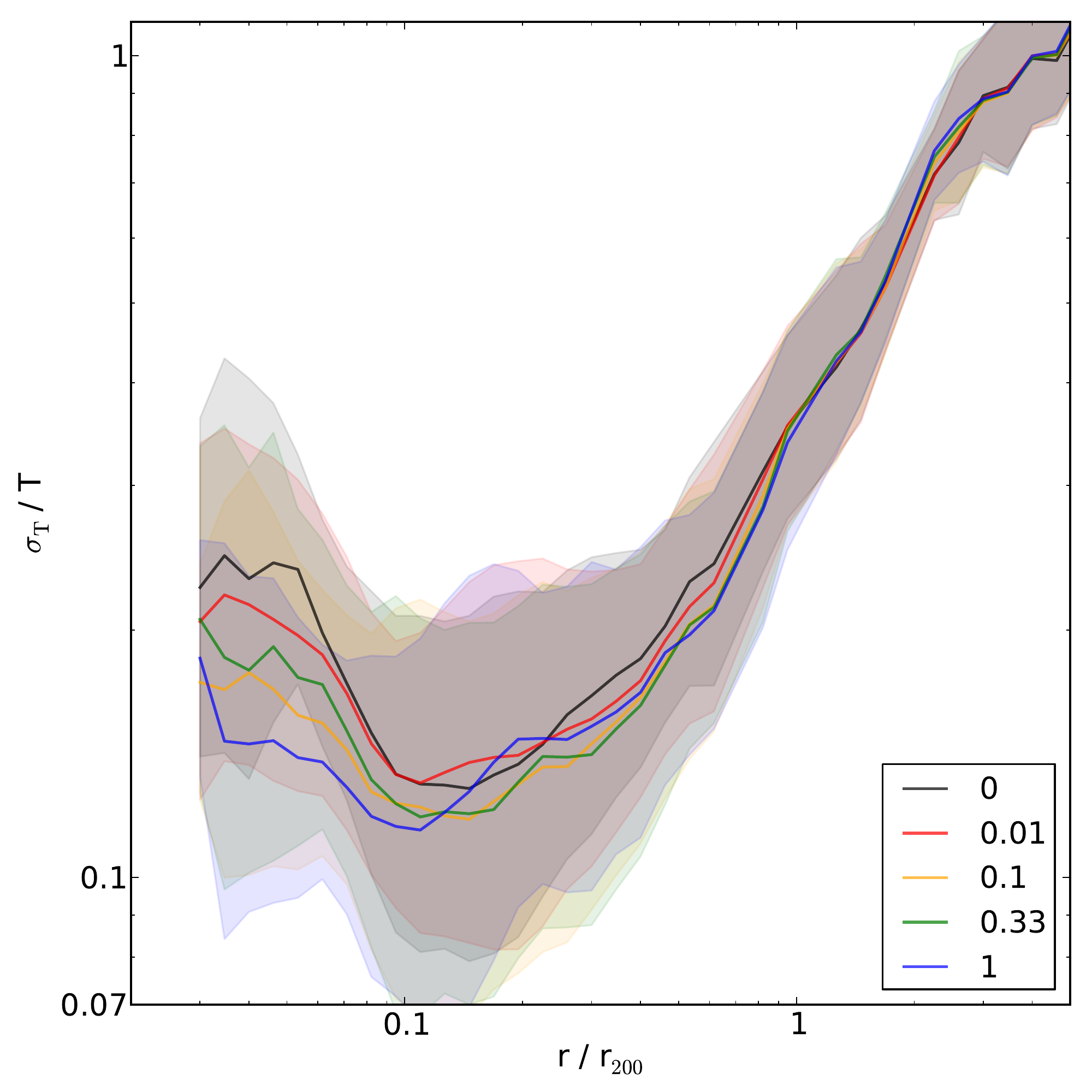}
  \caption{Averaged radial profile of the mass-weighted temperature
    variance normalized to mean temperature within each radial bin.  
    The process by which the averaged profiles are created is
    described in \S\ref{sec:profiles} and in the caption of Figure
    \ref{fig:radial_profiles}.
  } \label{fig:T_var}
\end{figure}

Conduction strong enough to alter the temperature gradients in
cluster cores should also be effective at smoothing out
small-scale thermal variations in the ICM.  However, those effects
turn out to be rather subtle, as shown in Figure \ref{fig:T_var}, which plots 
the normalized variance of the temperature field as a function of 
radius, averaged over all clusters in the sample that have the same 
level of conductivity.  It reveals an extremely weak trend of
greater temperature homogeneity with increasing $\fsp$, 
but at all radii the difference in homogeneity among cluster sets
with different values of $\fsp$ is less than the cluster-to-cluster 
variation.  Furthermore, this trend reverses beyond the virial radius, 
at $2 \lesssim \rnorm \lesssim 3$, albeit at a nearly marginal level.  

\begin{figure}
  \centering
  \includegraphics[width=0.45\textwidth]{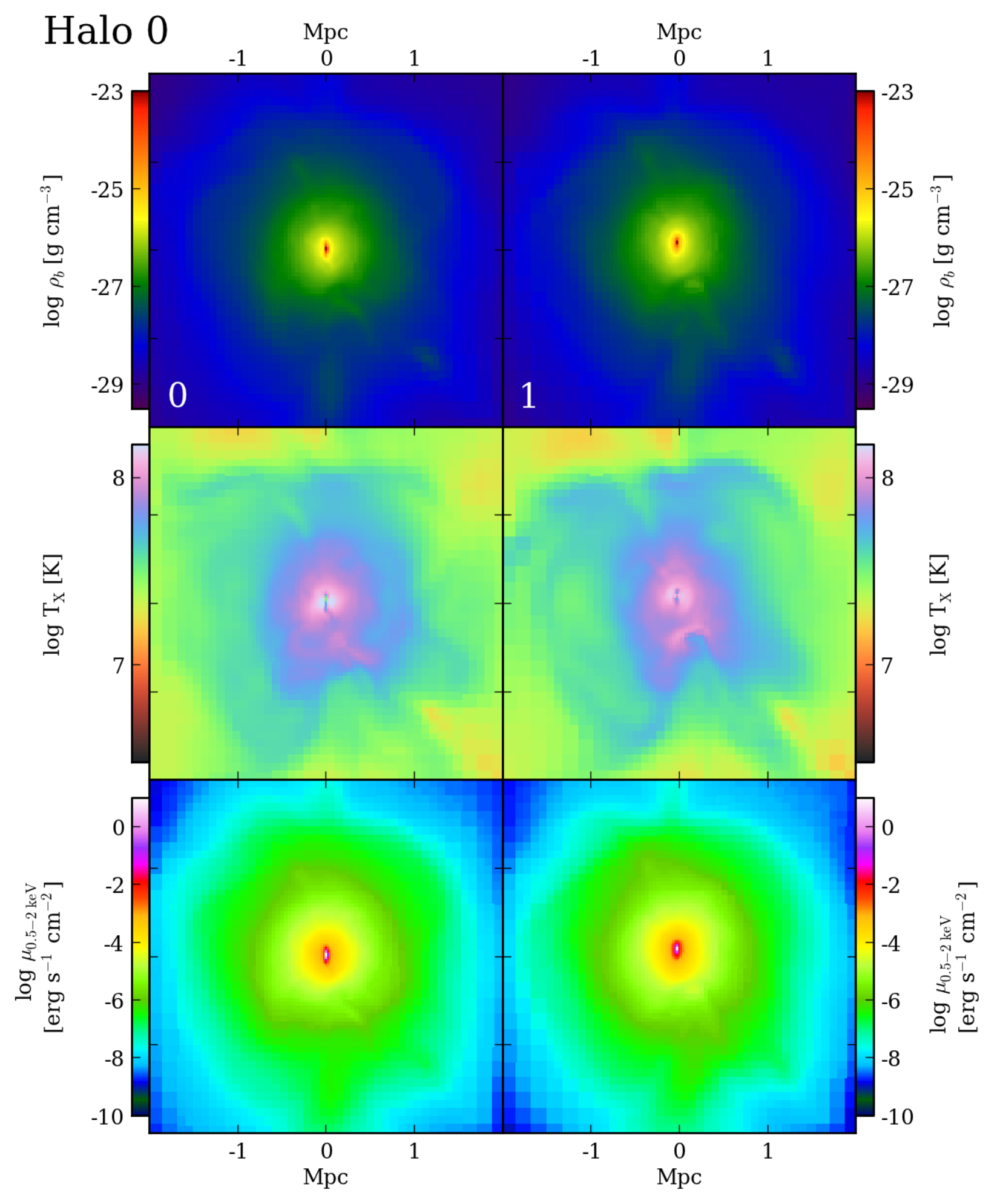}
  \caption{Similar projections to Figure \ref{fig:thin_proj_0} for
    Halo 0, but
    with a projected depth of 4 Mpc.
  } \label{fig:proj_0}
\end{figure}

\begin{figure}
  \centering
  \includegraphics[width=0.45\textwidth]{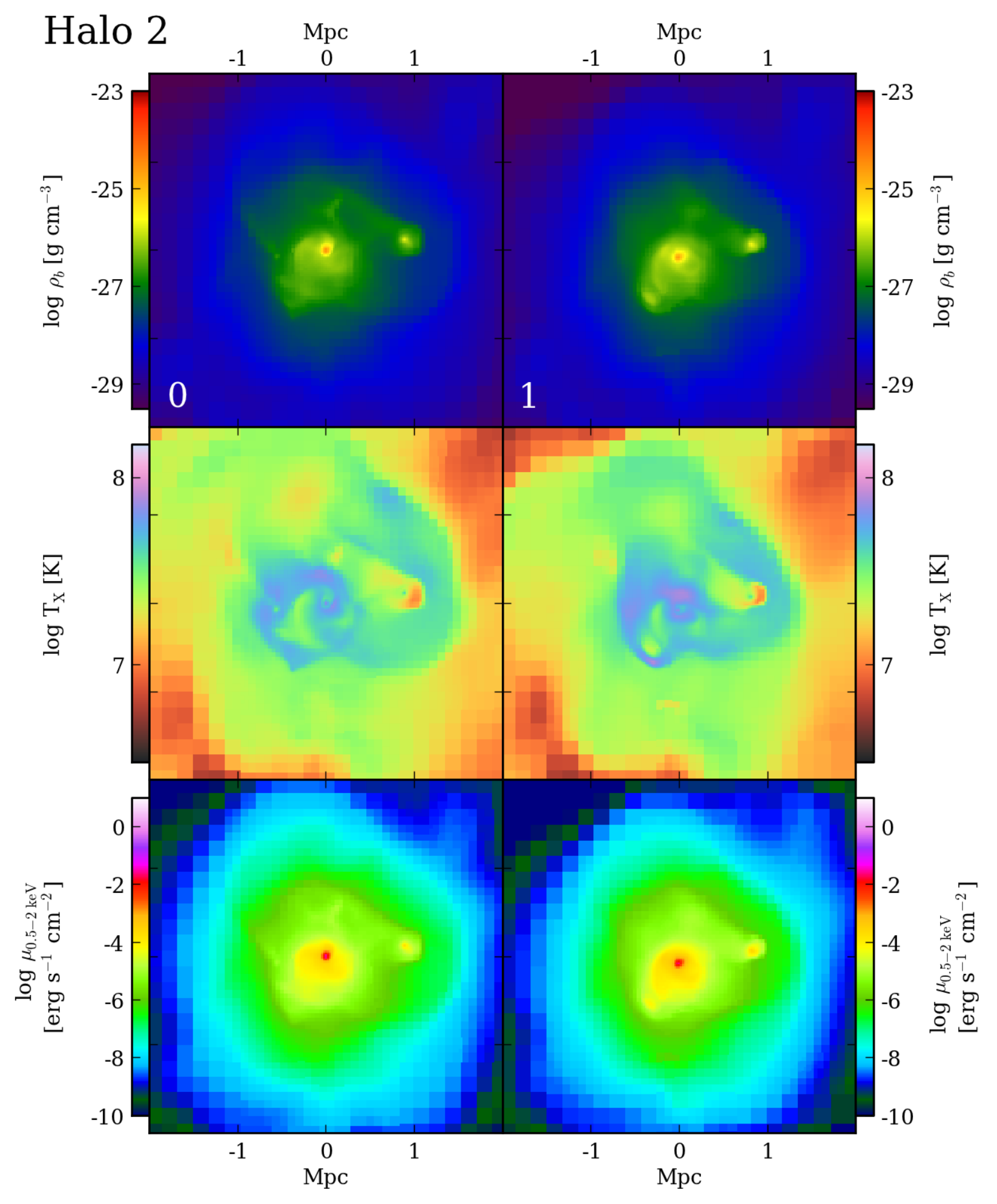}
  \caption{Similar projections to Figure \ref{fig:thin_proj_2} for
    Halo 2, but
    with a projected depth of 4 Mpc.
  } \label{fig:proj_2}
\end{figure}

\begin{figure}
  \centering
  \includegraphics[width=0.45\textwidth]{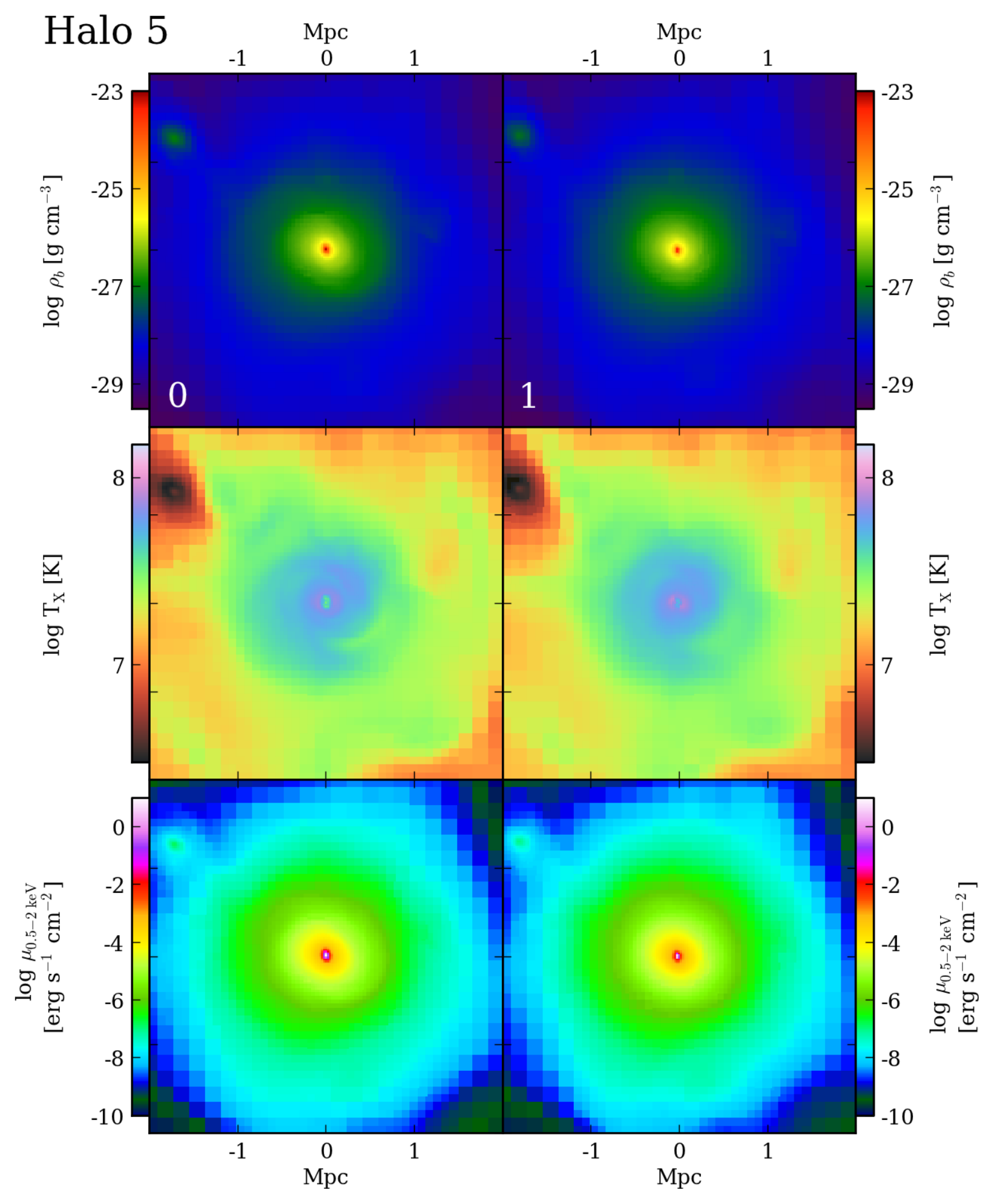}
  \caption{Similar projections to Figure \ref{fig:thin_proj_5} for
    Halo 5, but
    with a projected depth of 4 Mpc.
  } \label{fig:proj_5}
\end{figure}

The homogenizing effects of conduction are even harder to see in projection,
which perhaps is not surprising given the scarcity of obvious conduction-dependent
morphological differences in Figures \ref{fig:thin_proj_0}, \ref{fig:thin_proj_2}, and
\ref{fig:thin_proj_5}.  Those differences are further diluted when projected
over a 4~Mpc line of sight through each cluster,
as in Figures \ref{fig:proj_0}-\ref{fig:proj_5}. 
These latter figures show mean temperature weighted by the X-ray
emission in the 0.5-2 keV energy band, and X-ray emission is
calculated by interpolating from density, temperature, and
metallicity-dependent emissivity tables computed with the \texttt{Cloudy} code.  
Since X-ray emission is proportional to $n_e^{2} T_e^{1/2}$, this weighting
should highlight clumpiness and differences in temperature.  However, 
as stated previously, these results contrast considerably with the cluster map
comparison of \citet{2004ApJ...606L..97D}, which shows a significantly hotter 
cluster in which conduction should be much more efficient. 

\begin{figure}
  \centering
  \includegraphics[width=0.45\textwidth]{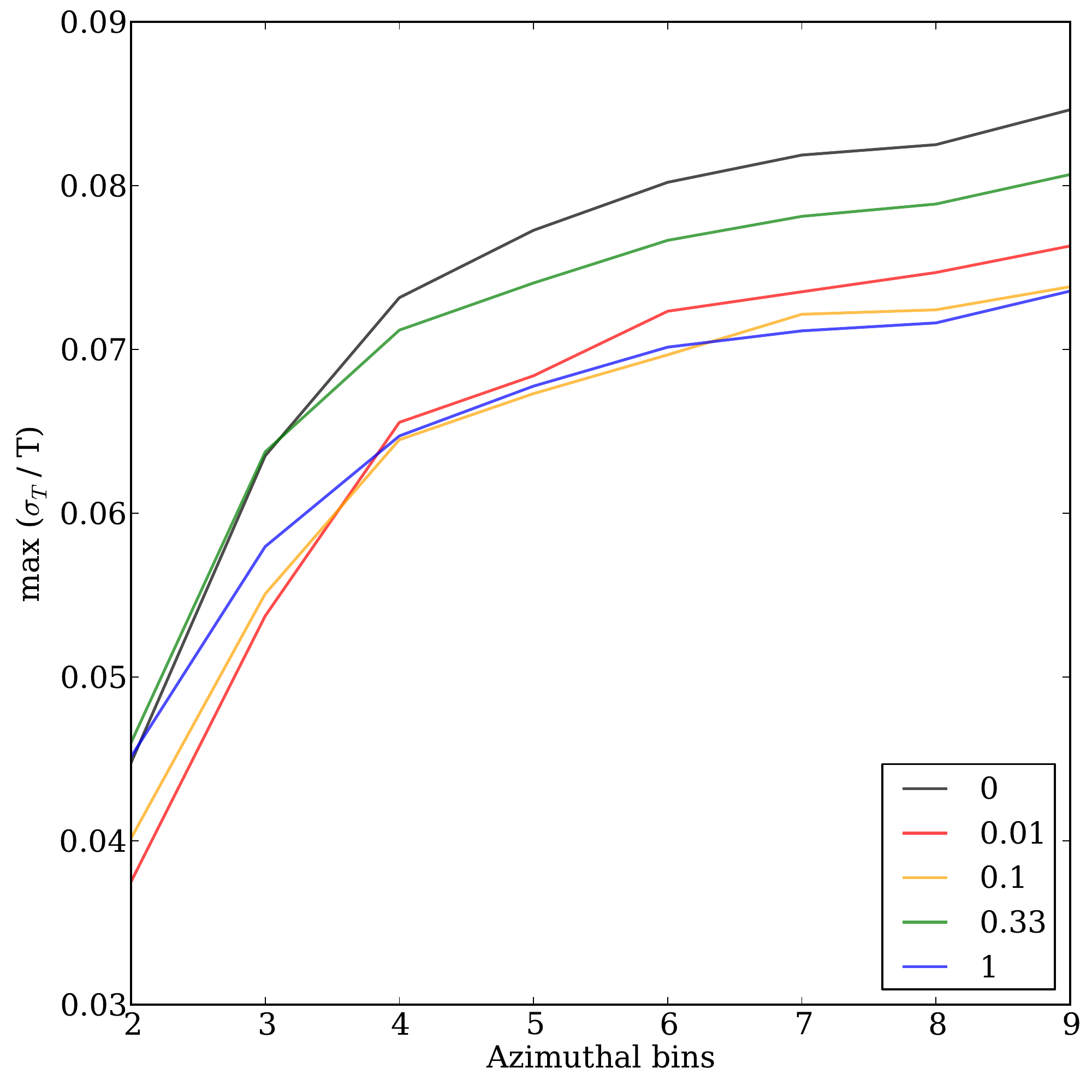}
  \caption{The maximum variance in X-ray-weighted temperature
    projections within a region 40 kpc $\le r \le$ 150 kpc as a
    function of the number of azimuthal bins.
  } \label{fig:azimuthal_variance}
\end{figure}

The change in temperature homogeneity due to conduction is quite
small in our simulated cluster sample, but important 
insights into the physics of the ICM could be gained if
the effective value of $\fsp$ could be measured observationally.  To
evaluate this possibility, we created X-ray-weighted temperature 
maps of the central 300 kpc for each of the clusters in the sample, masking
out the pixels within 40 kpc of the cluster centers to remove features 
that would be considered part of the central galaxy.  We then quantified the
amount of azimuthal temperature structure by dividing each 
temperature map into azimuthal bins and calculating both the mean 
temperature in each bin and the temperature variance among all azimuthal bins in
the map.  We performed this calculation multiple times for each map while rotating
the azimuthal bins, and took the maximum variance calculated as the 
value for that map.  Finally, we averaged the values together for all
clusters in the sample and performed the entire exercise over a range in
the total number of azimuthal bins, from 2 to 9.  Figure
\ref{fig:azimuthal_variance} plots the maximum variance as a
function of the number of bins for each value of $\fsp$.  We find
that, in general, the maximum variance is lower for the simulations
with conduction, but only by approximately 10\%.  There does
not appear to be any sort of monotonic trend with increasing $\fsp$.  We 
repeated this experiment, varying the inner and outer radius for the
temperature maps, but were unable to find conditions that produce a
better trend than can be seen in Figure \ref{fig:azimuthal_variance}.  
Thus, we conclude that the efficiency of conduction is difficult to
determine solely from the observable temperature homogeneity of the ICM,
at least for clusters of temperature $\lesssim 6$~keV.

\section{Star Formation}

\begin{figure}
  \centering
  \includegraphics[width=0.45\textwidth]{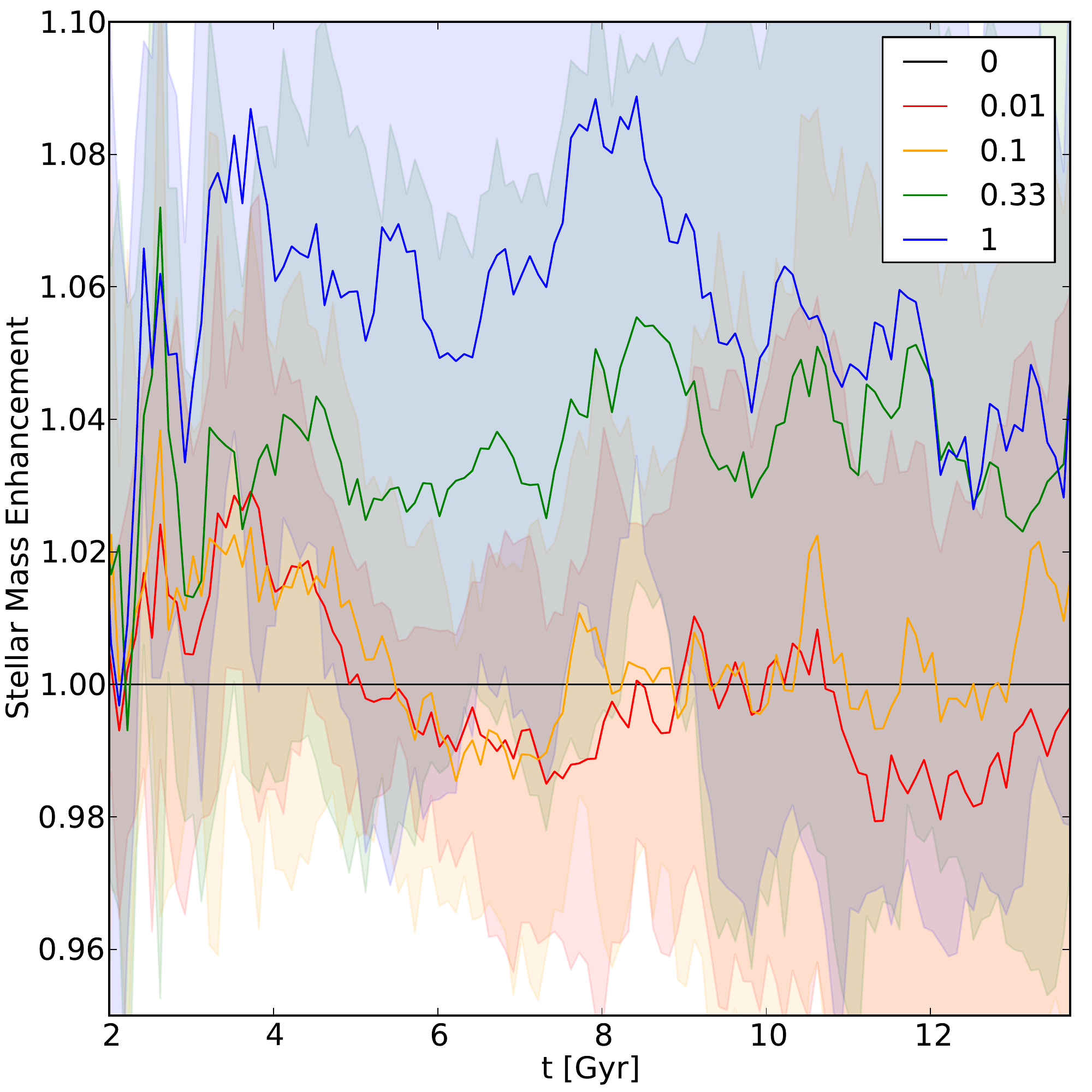}
  \caption{The ratio of total stellar mass in the simulations with
    $\fsp > 0$ to the control run as a function of cosmic time for
    all clusters in the sample, with the shaded regions denoting the
    cluster to cluster  variance.  The process by which the averaged
    quantities are calculated is described in \S\ref{sec:profiles} and
    in the caption of Figure \ref{fig:radial_profiles}.
  } \label{fig:stellar_mass}
\end{figure}

Figure \ref{fig:stellar_mass} shows the average difference in
stellar mass between the clusters with conduction and those without as
a function of time.  The difference in stellar mass at $z = 0$ for all
levels of conduction is less than 5\%.  Surprisingly, 
the clusters with higher levels of conduction form 
{\em more} stars in our calculations \citep[see also][]{2004ApJ...606L..97D}.  
However, the shaded
regions in Figure \ref{fig:stellar_mass} give an indication of just
how much variation there is between clusters.  As we have shown, even
the highest level of conduction is unable to prevent a cooling
catastrophe in the center of a cluster.  Therefore, one should not
expect a large difference in the amount of star formation.  Given the
coarseness with which the star forming regions are resolved, it is
possible that the enhancement in star formation with increasing $\fsp$
is numerical and not physical.  We propose two potential numerical explanations.  
First, the short timesteps required for the stability of the 
conduction algorithm may not provide enough time for a single star
particle to sufficiently heat up a grid cell and quench star formation
in a given cell in the following timestep.
Second, conduction may transport thermal energy too quickly out of
regions heated by recent star formation.  This would allow a 
star-forming region to recool and form additional stars too rapidly.  

Because conduction is able to create a nearly isothermal core for 
$\fsp \ge 0.1$, yet the final stellar mass from the clusters with $\fsp = 0.1$ 
is consistent with no change, we find it reasonable to conclude that 
the isothermality of the core in a conductive cluster has no influence 
on the star formation rate \texttt{Enzo} calculates for the central
galaxy.  However, because we do not resolve the interface between the
ICM and the interstellar medium (ISM), we cannot definitively state
the effect of conduction on star-forming gas in a cosmological simulation.

\section{Conclusions}

We have performed cosmological simulations of 10 galaxy clusters using
isotropic thermal conduction with five values of the conductive suppression
factor in order to study the effects of conduction on galaxy cluster
cores and the intracluster medium.  By studying the aggregate properties of the clusters in our
sample, we find that the presence of conduction even at
its maximum possible efficiency induces changes to
the density and temperature structure on the order of only 20-30\%.  For
$\fsp \ge 0.1$, the cluster cores become roughly isothermal.  However, 
conduction at any level is incapable of stopping the cooling
catastrophe at the very centers of our clusters, where the density profile
is always very sharply peaked.  To some extent, this is due to our limited
spatial resolution, since the temperature gradients on which 
heat conduction depends are limited by the scale of the smallest 
grid cell, which at $\sim15$ kpc/$h$ is still quite large compared 
to the scales of galaxies and the Field length.  

However, the extremely well-resolved study of
\citet{2012ApJ...747...26L} also finds that conduction can at best
slightly delay the cooling catastrophe.  While conduction is unable to
prevent the cooling catastrophe, the elevation of gas entropy in a conductive,
isothermal core displaces some of the core gas, 
moving it out to larger radii.  For values
of $\fsp$ up to 0.33, this material is redistributed mostly within
$\sim0.2~\rvir$.  For higher values of $\fsp$, it is transported out
even further, up to $\sim0.6~\rvir$.  A similar phenomenon occurs
around $\rvir$, where the negative temperature gradient allows outward
heat conduction to inflate the outer parts of the cluster.  
However, because this material is not deep in
the potential well, it is free to expand and cool, leading to slightly
lower temperatures just outside the virial radius.

More surprisingly, we observe a systematic decrease in both the
density and temperature with increasing $\fsp$ at large radii, out to
$\sim3~\rvir$.  We hypothesize that this is due to alteration of 
the accretion shocks by conduction.  To test this, we perform 
one-dimensional ``shock tube'' simulations with conditions characterizing
an accretion shock around a galaxy cluster, with the level of
conduction treated as the sole free parameter.
As long as the post-shock temperature is high enough for the Spitzer
conductivity to be efficient ($T \gtrsim 10^{7}$ K), conduction
moves the temperature jump upstream and the density jump downstream of
the original shock face.  This creates two distinct shocks, both with
Mach numbers less than the original shock.  We conclude that conduction is
responsible for the systematic decrease in density and temperature in
the outskirts of our simulated clusters, because it acts to weaken the shocks.  We
acknowledge that our modeling of this problem is not totally accurate,
and instead requires a two-fluid MHD approach, which is beyond the current
capabilities of our simulation tool.  However, more rigorous two-fluid simulations of
shockwaves in fully ionized plasmas show qualitatively similar
behavior, save a tiny feature in the ion temperature that cannot be
achieved in a single-fluid approach.  Unfortunately, because the effect on
clusters is only at the 10\% level and at very large radii, where the
X-ray surface brightness of the plasma is extremely low,
observing the effects of conduction on accretion shocks 
may never be possible.

We also find that in addition to altering temperature gradients,
conduction is able to make the intracluster medium more thermally uniform.
This effect, while measurable in spherically-averaged radial profiles,
is almost totally lost in projection.  Our results contrast with the
temperature maps of \citet{2004ApJ...606L..97D}, wherein the effect of
conduction is instantly recognizable.  
The cluster shown in \citet{2004ApJ...606L..97D} is
significantly more massive than our most massive cluster, so it is
possible that a hotter ICM, with a higher thermal conductivity, is
made more homogeneous, suggesting that the temperature dependence of
temperature inhomogeneity in a large cluster sample could help reveal the
typical conductivity of the ICM.  We attempted to find a means of distinguishing
the level of conduction observationally by measuring the variance in our
projected temperature maps, but without success.

Finally, conduction appears to have very little influence on the star formation 
rate within our simulated clusters.  When determining whether a grid
cell should form a star particle, we include the energy change from
conduction in the calculation of the cooling time, but this seems to
have very little influence.  This is likely because star-forming grid
cells are surrounded mostly by cells that are also quite cool.
Somewhat surprisingly, we observe a marginal increase in the
total stellar mass with increasing conduction, such that the sample with
$\fsp = 1$ shows an enhancement in star formation rate of $\sim5\%$.  The fact that
conduction cannot suppress star formation is directly related to its
inability to prevent the cooling catastrophe in the very center of the
cluster. However, the
reasons for the slight increase in star formation may be more
numerical than physical.  Further progress on understanding the 
effects of thermal conduction on star formation in cluster cores 
will require properly resolving the interface between ISM and the ICM, 
which at present is impractical in cosmological galaxy cluster simulations.

\acknowledgments

This work was supported by NASA through grant NNX09AD80G and
NNX12AC98G, and by the
NSF through AST grant 0908819.  The simulations presented here were 
performed and analyzed on the NICS Kraken and Nautilus supercomputing
resources under XSEDE allocations TG-AST090040 and TG-AST120009. 
We thank Greg Bryan, Gus Evrard, Eric Hallman, Andrey Kravtsov, Jack
Burns, Matthew Turk, and Stephen Skory for helpful discussions during
the course of preparing this 
manuscript.  SWS has been supported 
by a DOE Computational Science
Graduate Fellowship under grant number DE-FG02-97ER25308.  BWO was
supported in part by the MSU Institute for Cyber-Enabled Research.
\texttt{Enzo} and \texttt{yt} are developed by a large 
number of independent research from numerous institutions around the
world.  Their committment to open science has helped make this work
possible.


\end{document}